\documentclass[pra,aps,longbibliography,showpacs,groupedaddress,superscriptaddress,twocolumn,toc=flat,nofootinbib]{revtex4-1}

\usepackage{amsfonts,amsmath,amssymb,stmaryrd}

\usepackage[utf8]{inputenc}
\usepackage[T1]{fontenc}
\usepackage[table]{xcolor}
\usepackage{bbold, bm}
\usepackage{graphicx}
\usepackage{array,multirow}
\usepackage{boldline}
\usepackage{tabularx}
\usepackage{tikz}
\usepackage{placeins}
\usepackage{bm}
\usepackage{mathrsfs}
\usepackage{enumitem}   

\usepackage{ctable}
\usepackage{epsfig}
\usepackage[english, status=final]{fixme}
\fxsetup{marginface=\tiny}
\fxusetheme{color}    

\usepackage[colorlinks=true,linkcolor=blue,urlcolor=blue,citecolor=blue]{hyperref}

\newcommand{\bra}[1]{\langle#1|}
\newcommand{\ket}[1]{|#1\rangle}

\renewcommand{\H}{\hat{\mathcal{H}}}
\renewcommand{\c}{\hat{c}}
\newcommand{\cd}{\hat{c}^\dagger}

\newcommand{\h}{\hat{h}}
\newcommand{\hd}{\hat{h}^\dagger}

\newcommand{\p}{\hat{\pi}}
\newcommand{\pd}{\hat{\pi}^\dagger}
\newcommand{\s}{\hat{s}}
\newcommand{\sd}{\hat{s}^\dagger}
\renewcommand{\ij}{\langle \mathbf{i},\mathbf{j} \rangle}
\newcommand{\n}{\hat{n}}

\renewcommand{\j}{\mathbf{j}}
\newcommand{\xc}{\mathbf{x}_{\rm c}}

\AtBeginDocument{%
    \newwrite\bibnotes
    \def\bibnotesext{Notes.bib}
    \immediate\openout\bibnotes=\jobname\bibnotesext
    \immediate\write\bibnotes{@CONTROL{REVTEX41Control}}
    \immediate\write\bibnotes{@CONTROL{%
    apsrev41Control,author="08",editor="1",pages="1",title="0",year="1"}}
     \if@filesw
     \immediate\write\@auxout{\string\citation{apsrev41Control}}%
    \fi
}%

\begin{document}
\title{Scattering theory of mesons in doped antiferromagnetic Mott insulators: \\ Multichannel perspective and Feshbach resonance}

\author{Lukas Homeier}
\email{lukas.homeier@physik.uni-muenchen.de}
\affiliation{Department of Physics and Arnold Sommerfeld Center for Theoretical Physics (ASC), Ludwig-Maximilians-Universit\"at M\"unchen, Theresienstr. 37, M\"unchen D-80333, Germany}
\affiliation{Munich Center for Quantum Science and Technology (MCQST), Schellingstr. 4, M\"unchen D-80799, Germany}

\author{Pit Bermes}
\affiliation{Department of Physics and Arnold Sommerfeld Center for Theoretical Physics (ASC), Ludwig-Maximilians-Universit\"at M\"unchen, Theresienstr. 37, M\"unchen D-80333, Germany}
\affiliation{Munich Center for Quantum Science and Technology (MCQST), Schellingstr. 4, M\"unchen D-80799, Germany}

\author{Fabian Grusdt}
\email{fabian.grusdt@physik.uni-muenchen.de}
\affiliation{Department of Physics and Arnold Sommerfeld Center for Theoretical Physics (ASC), Ludwig-Maximilians-Universit\"at M\"unchen, Theresienstr. 37, M\"unchen D-80333, Germany}
\affiliation{Munich Center for Quantum Science and Technology (MCQST), Schellingstr. 4, M\"unchen D-80799, Germany}

\date{\today}

\begin{abstract}
Modeling the underlying pairing mechanism of charge carriers in strongly correlated electrons, starting from a microscopic theory, is among the central challenges of condensed-matter physics. Hereby, the key task is to understand what causes the appearance of superconductivity at comparatively high temperatures upon hole doping an antiferromagnetic (AFM) Mott insulator. Recently, it has been proposed~\cite{Homeier2023Feshbach} that at strong coupling and low doping, the fundamental one- and two-hole meson-type constituents -- magnetic polarons and bipolaronic pairs -- likely realize an emergent Feshbach resonance producing near-resonant $d_{x^2-y^2}$ interactions between charge carriers. Here, we provide detailed calculations of the proposed scenario by describing the open and closed meson scattering channels in the \mbox{$t$-$t'$-$J$~model} using a truncated basis method. After integrating out the closed channel constituted by bipolaronic pairs, we find \mbox{$d_{x^2-y^2}$} attractive interactions between open channel magnetic polarons. The closed form of the derived interactions allows us analyze the resonant pairing interactions and we find enhanced (suppressed) attraction for hole (electron) doping in our model. The formalism we introduce provides a framework to analyze the implications of a possible Feshbach scenario, e.g. in the context of BEC-BCS crossover, and establishes a foundation to test quantitative aspects of the proposed Feshbach pairing mechanisms in doped antiferromagnets.
\end{abstract}
\maketitle

%%%%%%%%%%%%%%%%%%%%%%%%%%%%%%%%%%%%%%%%%%%%%%%%%%%%%
\section{Introduction}
\label{sec:Introduction}
%%%%%%%%%%%%%%%%%%%%%%%%%%%%%%%%%%%%%%%%%%%%%%%%%%%%%

When the kinetic motion of constituents in a material is constrained by dominant repulsive Coulomb interactions, strongly correlated phases of electrons emerge, as observed in heavy fermions~\cite{Wirth2016}, Moir{\'e} materials~\cite{Andrei2021}, or quantum simulators~\cite{Bohrdt2021Review}, to name a few.
Another prominent example are underdoped cuprate compounds~\cite{Bednorz1986} with their numerous competing phases at low temperature~\cite{Lee2006,Fradkin2015,Keimer2015,Proust2019}, including the pseudogap phase, spin and charge order and superconductivity, which can be found in the phase diagram in the vicinity of an antiferromagnetic (AFM) Mott insulator.

The AFM Mott insulator constitutes a natural starting point to study the various phases that appear at low to intermediate doping~\cite{Lee2006}.
In this regime, the fundamental charge excitations are not the bare or renormalized electrons but have magnetic (or spin) polaron character~\cite{Badoux2016}, exhibiting strong correlations between the spin and charge degrees-of-freedom.
The magnetic polarons' properties have been quantitatively studied in great detail theoretically~\cite{Bulaevski1968,Brinkman1970,Trugman1988,Kane1989,Sachdev1989,Brink1998} and in numerical simulations~\cite{Trugman1990,Szczepanski1990,Chen1990, Johnson1991,Poilblanc1993,Poilblanc1993a,Dagotto1990,Liu1991}.
Additionally, ultracold atom quantum simulators with their direct access to spatial correlations provide an experimental platform to study the Fermi-Hubbard model, including the properties of individual magnetic polarons~\cite{Bohrdt2021Review}. This has enabled a direct observation of characteristic spin-charge correlations in a single magnetic polaron~\cite{Koepsell2019}, and similar polaronic features were observed up to dopings around $20\,\%$~\cite{Chiu2019} indicating an extended regime governed by magnetic polaron formation.

In parallel, a phenomenological theory of magnetic polarons was developed~\cite{Beran1996,Grusdt2018,Grusdt2019} which describes them as being composed of two partons carrying spin and charge quantum numbers, respectively, and which are confined into a mesonic bound state. This meson picture suggests that the magnetic polaron is not described by a dopant dressed with a featureless cloud of magnons, but rather the quasiparticle acquires a rich internal structure~\cite{Grusdt2018,Bohrdt2021_PRL} due to a rigid, confining string object~\cite{Bulaevski1968,Brinkman1970,Trugman1988,Manousakis2007,Grusdt2018,Grusdt2019} connecting the partons, i.e. the spinon~(s) and the chargon~(c).
We denote these mesons by their parton content as spinon-chargon~(sc). Indirect experimental evidence for the meson nature of the charge carriers below around $20\,\%$~doping has been obtained through the observation of string patterns~\cite{Chiu2019} and by machine-learning analysis of experimental data from cold atom quantum simulators~\cite{Bohrdt2019}. These studies suggest that string-formation itself plays an important role in the breakdown of AFM order as doping increases.

This meson picture is in stark contrast to earlier proposals by Anderson~\cite{Anderson1987}: He suggested that the electron (or hole) dopants fractionalize into free, i.e. deconfined, spinons~(s) and chargons~(c) in the finite doping regime.
This idea led to the development of emergent gauge theories with (de)confined partons~\cite{Lee1992,Senthil2000} and the resonating valence bond~(RVB) picture of unconventional superconductivity~\cite{Anderson1987,Lee2006}.
Today, the RVB picture of fractionalized partons remains debated however~\cite{Schrieffer2007}.

\begin{figure}[t!]
\includegraphics[width=\linewidth]{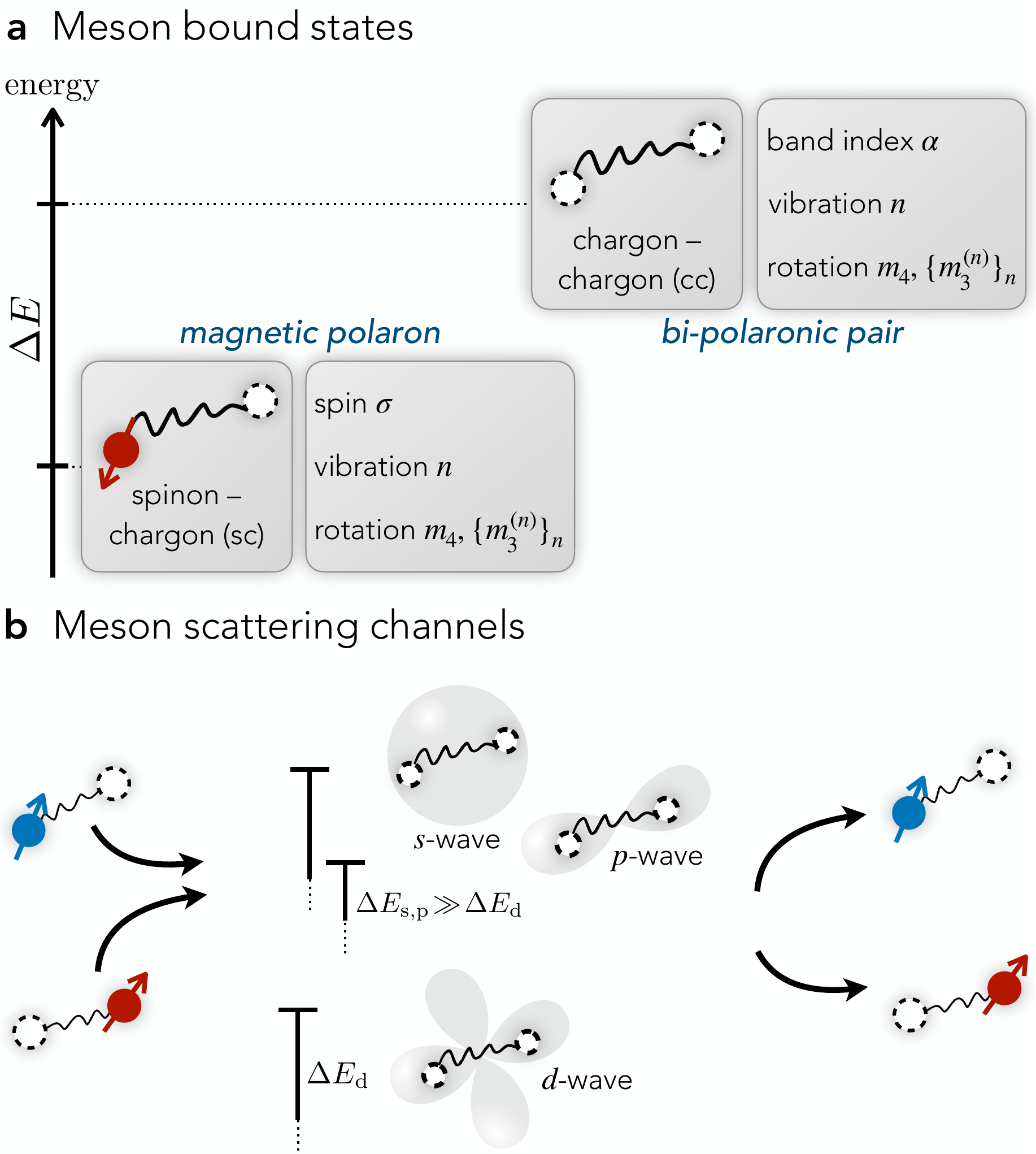}
\caption{\textbf{Mesons in doped antiferromagnets.} Understanding the properties of charge carriers in underdoped cuprates is essential to build microscopic, strong coupling theories of the various observed phases including the $d$-wave superconductor. Our starting point is the very low doping regime, i.e. one or two dopants in a strongly correlated AFM Mott insulator, which feature quasiparticles with rich internal structure. \textbf{a} The confined nature of partons, i.e. the spinon~(s) and chargon~(c), gives rise to mesonic~(sc) and (cc)~bound states. The mesons carry quantum numbers associated with their internal ro-vibration structure. \textbf{b} The Feshbach scattering scenario proposed in Ref.~\cite{Homeier2023Feshbach} describes two fermionic (sc)~mesons, which scatter via recombination processes into the bosonic, tightly-bound (cc)~state. The internal structure of the charge carriers leads to various scattering channels. If one channel approaches the scattering threshold, however, the scattering length diverges and dominates the low-energy physics. The Feshbach hypothesis of high-Tc superconductivity conjectures that cuprates remain on the BCS-side but are in close proximity to a \mbox{$d$-wave} resonant (cc)~state. }
\label{fig1}
\end{figure}

Instead, for a sufficiently long-ranged AFM correlation length~$\xi_{\rm{AFM}} \gtrsim a$, where~$a$ is the lattice spacing, the picture of confined partons discussed earlier is believed to resemble the situation in cuprates and Fermi-Hubbard type models. Here, the meson-like spinon-chargon (sc) bound states have the same quantum numbers as the electrons~\cite{Chowdhury2015} supplemented by a theoretically and numerically predicted rich internal structure~\cite{Brinkman1970,Shraiman1988,Trugman1988,Simons1990,Beran1996,Brunner2000,Mishchenko2001,Bohrdt2020}.
Additionally, recent density matrix renormalization group (DMRG) simulations~\cite{Bohrdt2023_dichotomy} and analytical studies~\cite{Grusdt2023} find light, long-lived two-hole resonances~\cite{Poilblanc1994} described by a bosonic chargon-chargon (cc)~meson and with similar internal structure.
Solving for the parton bound state exactly is challenging, however phenomenological models have been put forward predicting a rich ro-vibrational string-like excitation spectrum~\cite{Trugman1988,Simons1990,Manousakis2007,Grusdt2018,Grusdt2019} of the fermionic magnetic polaron and the bosonic bi-polaron~\cite{Grusdt2023,Bohrdt2023_dichotomy,Shraiman1988}, see Fig.~\ref{fig1}a.

Motivated by experimental evidence in cuprate compounds -- indicating the fermionic character of charge carriers in the ground state~\cite{DoironLeyraud2007,Shen2005,Sous2023,Kurokawa2023} -- one approach is to formulate a low-doping and strong coupling model based on a normal state constituted by weakly interacting magnetic polarons; these describe the fermionic quasiparticles of individual mobile holes doped into an AFM Mott insulator. This approach is further justified by recent ARPES studies~\cite{Kunisada2020,Kurokawa2023} in the extremely low-doping regime of layered cuprates indicating the existence of a Fermi surface around the nodal points~$\mathbf{k}=(\pm\pi/2,\pm\pi/2)$~\cite{Shen2005}, which is consistent with the predicted hole pockets of magnetic polarons~\cite{Kane1989,Sachdev1989,Brink1998,Trugman1990,Poilblanc1993a,Dagotto1990}.
Experiments in this compound suggest that any small hole doping of the AFM Mott insulator gives rise to a metallic state with onset of superconductivity at a doping of around~$4\,\%$ seen by the opening of a pairing gap~\cite{Kurokawa2023}.

An exceptional effort has been put forward to explain the origin of the strong pairing in cuprates, starting from various proposed parent states including the magnetic polarons; hereby numerous studies have shown the importance of magnetic fluctuations for pairing~\cite{Miyake1986,Scalapino1986,Schrieffer1988,Su1988,Millis1990,Monthoux1991,Scalapino1995,Schmalian1998,Halboth2000,Abanov2003,Bruegger2006,Metzner2012,Vilardi2019}.
However, the charge carrier's strong coupling nature, i.e. their emergence from the underlying correlated background, prohibits to develop a simple interacting theory at finite doping.
In particular, developing a unifying description that includes the rich microscopic structure of the emergent charge carriers, i.e. the string and its fluctuations, has remained challenging. The goal of this article is to formulate a low but finite doping description of these charge carriers fully including their internal structure.

In an accompanying study~\cite{Homeier2023Feshbach}, the scenario of an emergent Feshbach resonance in underdoped cuprates is proposed.
It is argued that the (sc)'s, forming the normal state of a doped AFM insulator, can recombine with a light, bi-polaronic, near-resonant (cc)~channel leading to a scattering resonance.
Hereby, the scattering symmetry is dictated by the rotational symmetry of the tightly-bound (cc)~state giving rise to attractive \mbox{$d$-wave} interactions; in principle off-resonant scattering processes of a different symmetry can contribute weakly, see Fig.~\ref{fig1}b.
This provides a new perspective on the origin of pairing between charge carriers, i.e. magnetic polarons, mediated by a closed (cc)~scattering channel and based on magnetic interactions~\cite{OMahony2022}.
In Ref.~\cite{Homeier2023Feshbach} the Feshbach hypothesis of high-Tc superconductivity in cuprates is formulated, which conjectures that cuprate superconductors are in the vicinity of a \mbox{$d_{x^2-y^2}$-wave} scattering resonance, but on the BCS side where the ground state is constituted by magnetic polarons~(sc).

In this article, we collect further evidence for the Feshbach scattering scenario. To this end, we develop a truncated basis method to obtain the mesons by confining strings, and apply it to describe a multichannel model of magnetic polarons~(sc)$^2$ in the open channel and tightly-bound (cc)~states in the closed channel, see Sec.~\ref{sec:Model} and~\ref{sec:OpenClosed}.
In Sec.~\ref{sec:ScattLength}, we calculate the matrix elements of the resonant interactions by performing a controlled approximation in the length of strings and by considering open-closed channel recombination processes induced by spin-flip~$J_\perp$ and weak next-nearest neighbouring (NNN) tunneling~$t'$ events.
We compare our results in Sec.~\ref{sec:truncated-basis} to a more quantitative and refined truncated basis method, that cures the overcompletness of the strings states and allows us to systematically include non-perturbative effects associated with larger values of~$t'$.

As we show, after integrating out the near-resonant closed channel, the mediated low-energy scattering interactions give rise to \mbox{$d_{x^2-y^2}$-wave} BCS-type pairing between two magnetic polarons with zero total momentum, see Sec.~\ref{sec:effModel}. 
We discuss additional results and immediate consequences that follow from the Feshbach pairing mechanism obtained in the semi-analytical description of the meson bound states:
In the limit of weak NNN tunneling~$|t'| \ll |t|$, we predict stronger pairing interactions in hole-doped than electron-doped compounds.
Further, our model allows us to obtain a set of BCS mean-field equations predicting a \mbox{$d_{x^2-y^2}$-wave} pairing gap.
Lastly, we discuss experimental signatures and derive coupling matrix elements for single hole ARPES and coincidence ARPES in Sec.~\ref{sec:ARPES}.
We conclude with a summary and outlook in Sec.~\ref{sec:summary-outlook}.

%%%%%%%%%%%%%%%%%%%%%%%%%%%%%%%%%%%%%%%%%%%%%%%%%%%%%
\section{Model}
\label{sec:Model}
%%%%%%%%%%%%%%%%%%%%%%%%%%%%%%%%%%%%%%%%%%%%%%%%%%%%%

The starting point to describe one and two dopants (holes or electrons) in the AFM Mott insulator is a 2D square lattice $t$-$t'$-$J$~model, which is the low-energy description of the the strong coupling~$U \gg t$ Fermi-Hubbard model~\cite{Auerbach1994}.
The Hamiltonian of the model is given by
\begin{align}
\begin{split}\label{eq:tJ}
    \H_{\text{$t$-$t'$-$J$}} = &-t \sum_{\ij, \sigma} \hat{\mathcal{P}}\left( \cd_{\mathbf{i},\sigma}\c_{\mathbf{j},\sigma} +\mathrm{h.c.} \right)\hat{\mathcal{P}}\\
    &-t' \sum_{\langle\!\ij\!\rangle, \sigma}\hat{\mathcal{P}}\left( \cd_{\mathbf{i},\sigma}\c_{\mathbf{j},\sigma} +\mathrm{h.c.} \right)\hat{\mathcal{P}}\\
    &+ J_z \sum_{\ij} \left( \hat{S}^z_\mathbf{i}  \hat{S}^z_\mathbf{j}  - \frac{1}{4}\n_\mathbf{i} \n_\mathbf{j}  \right) \\
    &+ \frac{J_\perp}{2} \sum_{\ij} \left( \hat{S}^+_\mathbf{i} \hat{S}^-_\mathbf{j} +\mathrm{h.c.} \right),
\end{split}
\end{align}
where~$\c_{\mathbf{j},\sigma}$ describes the underlying electrons with spin~$\sigma=\downarrow,\uparrow$ at site~$\mathbf{j}$; the spin-1/2 operator~$\hat{\mathbf{S}}_\mathbf{j}=\frac{1}{2}\cd_{\mathbf{j},\sigma} \hat{\bm{\tau}}_{\sigma\sigma'} \c_{\mathbf{j},\sigma'}$ is constructed from Pauli matrices~$\bm{\tau}$.
Furthermore, the Gutzwiller projector ensures that the particle number, $\n_{\mathbf{j}}=\n_{\mathbf{j},\downarrow} + \n_{\mathbf{j},\uparrow}$,  is constrained to $\n_{\mathbf{j}} \leq 1$ for all sites~$\mathbf{j}$ akin to strong Hubbard repulsion in the parent model.
The first two terms in Hamiltonian~\eqref{eq:tJ} describe NN and NNN tunneling with amplitude~$t$ and~$t'$, respectively.
The last two terms are the effective AFM interaction with superexchange strength~$J_z =  J_\perp = +4t^2/U$.
In the strong-coupling limit -- typically $t/J \simeq 3$ is assumed in cuprate materials -- the undoped ($\n_{\j}=1$ for all~$\mathbf{j}$) ground state~$\ket{0}$ is AFM N\'eel ordered.

\begin{figure}[t!]
\includegraphics[width=\linewidth]{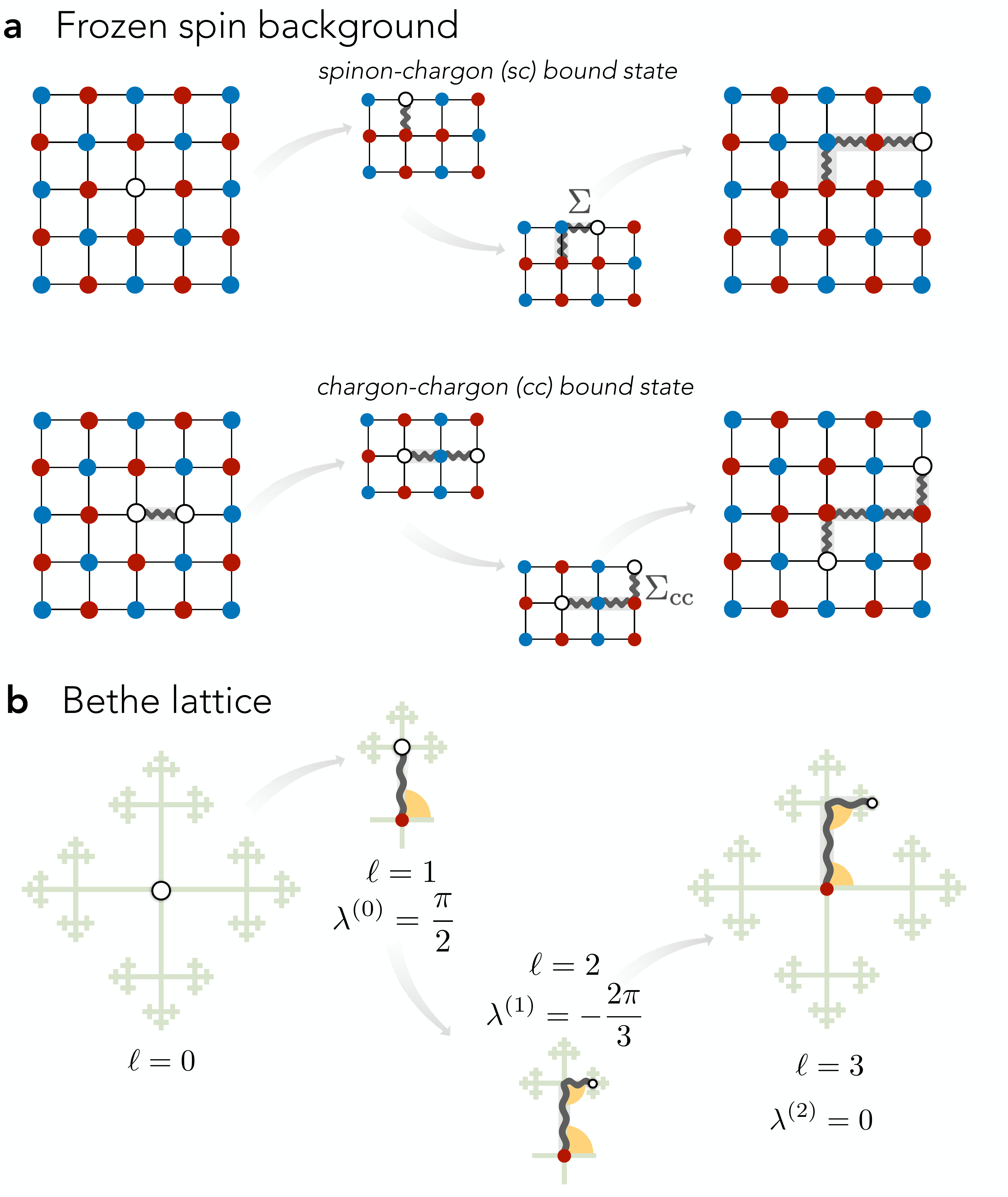}
\caption{\textbf{Truncated basis of geometric strings.} \textbf{a} We describe the single dopant (magnetic polaron) in a parton framework giving rise to a spinon-chargon bound state~(sc). For~$t \gg J$, the chargon moves through a frozen spin background, i.e. the latter adapts on much slower time scales,~$\tau_{\rm s} \propto J_\perp^{-1}$, then the former, $\tau_{\rm c} \propto t^{-1}$. The chargon's motion rearranges the spin background leaving a memory, which we encode in the geometric string~$\Sigma$ (gray curly line). The displacement of spins leads to parton confinement akin to a linear string tension~$\propto J_z$. Analogously, the parton construction is applicable to the two dopant problem, leading to a tightly-bound, bi-polaronic chargon-chargon (cc) bound state connected by a string~$\Sigma_{\rm cc}$ (bottom). \textbf{b} We map the string states on the sites of a Bethe lattice. Here, we illustrate the hopping events of the chargon as in~\textbf{a} (top). The rotational symmetries of the Bethe lattice allow us to assign well-defined rotational quantum numbers to the magnetic polaron for $C4$-invariant momenta. We associate an angle~$\lambda^{(N)}$ between string elements of length~$\ell=N$ and~$\ell=N+1$, where~$\ell$ is the length of the string.}
\label{figBethe}
\end{figure}

%%%%%%%%%%%%%%%%%%%%%%%%%%%%%%%%%%%%%%%%%%%%%%%%%%%%%
\section{Open and closed channel description}
\label{sec:OpenClosed}
%%%%%%%%%%%%%%%%%%%%%%%%%%%%%%%%%%%%%%%%%%%%%%%%%%%%%
In the following, we recap the geometric string formalism in order to describe mobile single dopant (sc) and two dopant (cc) impurities immersed into an AFM Mott insulator~$\ket{0}$; we closely follow Refs.~\cite{Grusdt2018,Grusdt2023}.
We review the basic concepts and introduce the notation required for the calculation of the Feshbach scattering length.

The geometric strings originate from the dopant's displacement of the AFM ordered spin background, see Fig.~\ref{figBethe}, and was pointed out in early theoretical studies of the Hubbard or \mbox{$t$-$J$}~model by Brinkman and Rice~\cite{Brinkman1970}, Trugman~\cite{Trugman1988} and Beran~et~al.~\cite{Beran1996}, among others.
The rigid string-like object naturally gives rise to a rich internal structure of the dopant's quasiparticle, explaining the long-lived vibrational excitations revealed in numerical~\cite{Beran1996,Manousakis2007,Brunner2000,Mishchenko2001,Bohrdt2021} and analytical studies~\cite{Bulaevski1968,Kane1989,Liu1991,Grusdt2018,Grusdt2019}.
The string picture does not only provide a phenomenological explanation of the spectral features, but it can also be put in a stringent quantitative formalism, i.e. the geometric string theory, based on a semi-analytical truncated basis approach.
In this approach, a variational wavefunction for the (sc) and (cc) bound states in the string basis is obtained.
The truncated basis we introduce below is an exact description for the (sc) and (cc) bound states in the $t$-$J_z$~limit ($J_\perp = t' = 0$).
However, numerical simulations indicate that the qualitative features of the string description remain valid in the \mbox{$t$-$J$}~model ($J_z=J_\perp$, $t'=0$)~\cite{Liu1991,Brunner2000,Trugman1990,Szczepanski1990,Chen1990,Johnson1991, Poilblanc1993,Poilblanc1993a,Dagotto1990,Liu1991,Grusdt2019,Bohrdt2020,Bohrdt2021_PRL,Bohrdt2021,Bohrdt2023_dichotomy}, and experiments of the Fermi-Hubbard model in ultracold atoms show direct~\cite{Koepsell2019} and indirect~\cite{Chiu2019,Bohrdt2019} signatures of string correlations.

\begin{figure}[t!]
\includegraphics[width=\linewidth]{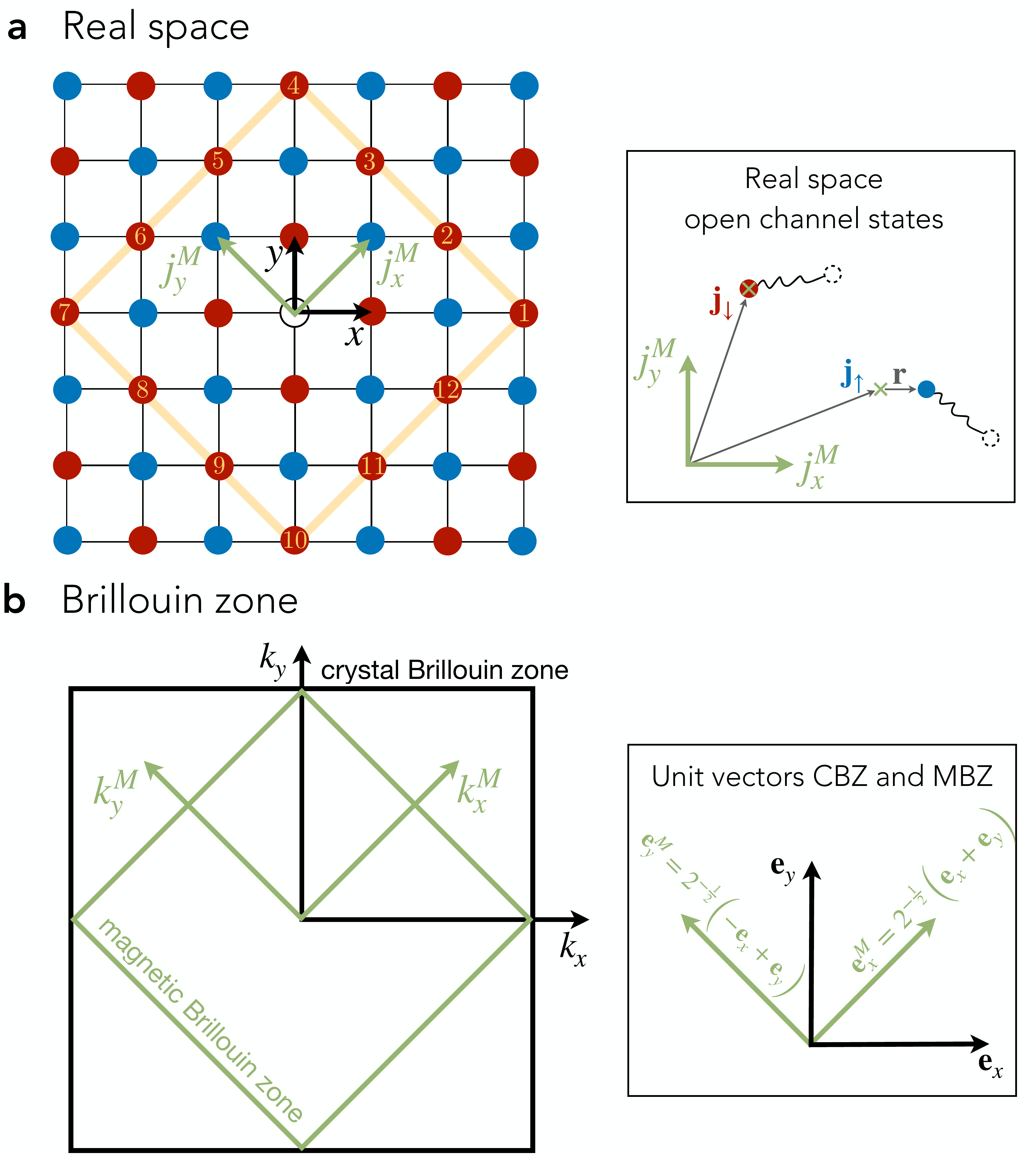}
\caption{\textbf{Crystal and magnetic lattice.} \textbf{a} The long-range magnetic spin order breaks translational symmetry of the underlying crystal lattice. The magnetic lattice has a two site unit cell with basis vectors~$\mathbf{j}_\downarrow =(j_x^M, j_y^M)$. \textbf{b} We illustrate the corresponding Brillouin zones of the crystal (CBZ) and magnetic (MBZ) Brillouin zone. The scattering calculations are performed in the MBZ with momenta~$\mathbf{k}^M=(k_x^M,k_y^M)$. }
\label{figMBZ}
\end{figure}
%%%%%%%%%%%%%%%%%%%%%%%%%%%%%%%%%%%%%%%%%%%%%%%%%%%%%
\subsection{Spinon-chargon (sc) bound state: \\ Magnetic polarons}
\label{sec:sc-stringtheory}
%%%%%%%%%%%%%%%%%%%%%%%%%%%%%%%%%%%%%%%%%%%%%%%%%%%%%
\emph{Individual magnetic polaron.---}
In the following, we consider a (classical) N\'eel state~$\ket{0}$ with long-range AFM order and a single mobile dopant, i.e. the magnetic polaron or (sc) bound state. The perfect N\'eel background should be a justified approximation when the correlation length~$\xi \gg a$ exceeds several lattice constants. To describe the (sc) bound state, we construct a Krylov basis by applying the hopping terms~$t$ in the Hamiltonian to the state~$\hat{c}_{\mathbf{j},\bar{\sigma}}\ket{0}$ leading to string states~$\Sigma$ discussed below.
Thus, the truncated basis for the (sc) bound state is spanned by~$\{ \ket{\mathbf{j}_\sigma,\Sigma} \}$ with spinon position~$\mathbf{j}_\sigma$ and string~$\Sigma$ that connects the spinon to a spinless chargon, see Fig.~\ref{figBethe}a.

Let us describe the physical origin of the string~$\Sigma$.
In the strong-coupling limit, $t \gg J$, the time scales of the dopant's motion, $\tau_{\rm c} \propto t^{-1}$, and magnetic background, $\tau_{\rm s} \propto J_\perp^{-1}$, decouple and we can treat the problem in Born-Oppenheimer approximation, i.e. we choose a product state ansatz for the (sc) bound state~$\ket{\psi_{\rm{sc}}}\approx\ket{\psi_{\rm{s}}}\otimes\ket{\psi_{\rm{c}}}$ by decomposing the wavefunction into its spinon~$\ket{\psi_{\rm{s}}}$ and chargon~$\ket{\psi_{\rm{c}}}$ contribution, see e.g. Ref.~\cite{Grusdt2018}.

To this end, we consider a single hole~$\c_{\mathbf{j},\bar{\sigma}}$ (electron~$\c^\dagger_{\mathbf{j},\sigma}$) doped into a N\'eel background~$\ket{0}$, creating a spinon at position~$\mathbf{j}_\sigma$.
The fast motion~$\propto t$ of the chargon distorts the magnetic order before the magnetic background can adapt on its intrinsic time scale~$\tau_{\rm s} \gg \tau_{\rm c}$.
Thus, in the so-called frozen spin approximation, we consider the motion of the chargon through a static background of spins, where the chargon's motion rearranges the spins; this gives rise to states that we label by~$\ket{\mathbf{j}_\sigma,\Sigma}$ (see Appendix).

We emphasize that the spinon and chargon position above are not sufficient to describe the state but one needs to take into account the chargon's path, i.e. the string~$\Sigma$, which begins at the spinon position~$\mathbf{j}_\sigma$ and ends at the chargon's position, see Fig.~\ref{figBethe}a.
Even in a perfect N{\'e}el background the string states $\{ \ket{\mathbf{j}_\sigma,\Sigma} \}$ have an overcompleteness originating from so-called Trugman loops~\cite{Trugman1988}, where some spin configurations can be described by multiple string states.
The effect of Trugman loops has been shown to be subdominant~\cite{Grusdt2018} to capture the chargon wavefunction~$\ket{\psi_{\rm c}}$ but is important to describe the fine features with precision of a fraction of~$J$ in the magnetic polaron's dispersion~\cite{Bermes2024}.
In Sec.~\ref{sec:truncated-basis} we will treat loop effects systematically in this model; for now we follow Ref.~\cite{Grusdt2018} and assume $\{ \ket{\mathbf{j}_\sigma,\Sigma} \}$ to form an orthonormal basis set, for which $\ket{\psi_{\rm{c}}}$ can be determined by solving a single-particle problem on the Bethe lattice, as we describe next.

In the orthonormal basis set, the string states can be uniquely characterized by (i) their length~$\ell$, which is the depth on the Bethe lattice, and (ii) the angle~$\lambda^{(N)}$ between the $N$-th and $(N+1)$-th string element, see Fig.~\ref{figBethe}b.
This allows us to re-label the string states
\begin{align}
    \ket{\mathbf{j}_\sigma, \Sigma} = \ket{\mathbf{j}_\sigma, \ell, \lambda^{(0)} , \lambda^{(1)},...},
\end{align}
where~$\lambda^{(0)} = 0,\,\pi/2,\,\pi,\,3\pi/2$ and $\lambda^{(N)} = -2\pi/3,\,0,\,2\pi/3$ for~$N>0$.
Note, however, that the translational invariance as well as the C4-invariance of the square lattice model can only be simultaneously exploited at $C4$-invariant momenta, i.e. at momenta~$\mathbf{k}^M=(0,0)$ and $\mathbf{k}^M =(\pi/\sqrt{2},\pi/\sqrt{2})$ in the magnetic Brillouin zone (MBZ).

The MBZ is defined as follows.
Since the N\'eel AFM breaks the sublattice symmetry, the spinon position~$\mathbf{j}_\sigma$ is defined in the doubled, AFM unit cell.
As a consequence, momenta~$\mathbf{k}^M$ are formally defined in the MBZ with band index~$\sigma=\downarrow, \uparrow$, which is obtained by reducing the volume of the crystal Brillouin zone (CBZ) by~$1/2$ and by rotating the CBZ by~$\pi/4$, see Fig.~\ref{figMBZ}.
To be precise, the unit vectors in the MBZ are given by
\begin{align} \label{eq:unitvecs_MBZ}
\mathbf{e}_x^M = \frac{1}{\sqrt{2}}\left(  \mathbf{e}_x + \mathbf{e}_y \right) ~~~~~ \mathbf{e}_y^M = \frac{1}{\sqrt{2}}\left( - \mathbf{e}_x + \mathbf{e}_y \right)
\end{align}
such that momentum vectors in the first MBZ are given by~$\mathbf{k}^M = k_x^M \mathbf{e}_x^M + k_y^M \mathbf{e}_y^M$ with $k_x^M, k_y^M \in \left[ -\frac{\pi}{\sqrt{2}},\frac{\pi}{\sqrt{2}} \right]$.
The MBZ momenta~$\mathbf{k}^M$ can always be obtained by folding momenta~$\mathbf{k}$ from the CBZ into the MBZ; hence we only use the superscript if needed.

Next, we evaluate the \mbox{$t$-$J$}~Hamiltonian in the string basis, where the hopping term connects different string states, the Ising term~$J_z$ corresponds to a confining potential~$\propto \ell$, and the flip-flop terms~$J_\perp$ give rise to spinon dispersion (see Appendix).
At $C4$-invariant momenta, the total momentum~$\mathbf{k}$ and rotational eigenvalues~$\{ m_4, m_3^{(1)},m_3^{(2)},... \}$ with~$m_4=0,...,3$ ($s$-, $p$-, $d$- and $f$-wave) and $m_3^{(N)}=0,...,2$ form a set of quantum numbers for the magnetic polaron, see Fig.~\ref{fig1}, and the wavefunction of the $n$-th eigenstate can be written as~\cite{Grusdt2018}
\begin{widetext}
\begin{align}
\begin{split}\label{eq:sc_boundstates}
    \ket{\mathbf{k}, \sigma, n, m_4, \{ m_3 \}} = \left(\frac{L^2}{2}\right)^{-1/2} \sum_{\mathbf{j}_\sigma} e^{-i\mathbf{k}\mathbf{j}_\sigma} \sum_{\ell} \sum_{M=0}^{\ell-1}\sum_{\lambda^{(0)}}...\sum_{\lambda^{(M)}} e^{-i \left[ \lambda^{(0)} m_4 + \sum_{N=1}^{M}\lambda^{(N)}m_3^{(N)}\right] } \times \\
    \times \frac{1}{\sqrt{4}}\left(\frac{1}{\sqrt{3}} \right)^M \psi_{\rm{sc}}^{(n)}(\mathbf{k},\ell,m_4,\{m^{(M)}_3\})\ket{\mathbf{j}_\sigma, \ell, \{\lambda^{(M)} \} }.
\end{split}
\end{align}
\end{widetext}
The spinon's spin quantum number~$\sigma$ defines the sublattice of the magnetic polaron and $L^2$ is the volume of the underlying crystal lattice.
Moreover, $\psi_{\rm{sc}}^{(n)}(\mathbf{k},\ell,m_4,\{m^{(N)}_3\})\in\mathbb{R}^{\geq 0}$ are the amplitudes of the normalized wavefunction, which depend on the total momentum~$\mathbf{k}$ and the internal degrees-of-freedom; note that we have made a choice of gauge for the string states~$\{ \ket{\mathbf{j}_\sigma,\Sigma} \}$ in order to obtain positive and real wavefunction amplitudes (see Supplementary Information).
Away from the $C4$-invariant momenta, the angular momentum is not a good quantum anymore and the different sectors~$\{ m_4, m_3^{(1)},m_3^{(2)},... \}$ hybridize.

This variational approach agrees with full numerical calculations~\cite{Brunner2000,Mishchenko2001,Bohrdt2020}, captures the scaling of the ground-state energy~$\propto t^{1/3}J^{2/3}$, and explains the (gapped) excitation spectrum in terms of ro-vibrational string excitations~\cite{Simons1990,Bohrdt2021}.
The magnetic polaron has minimal energy at the nodal points~$\mathbf{k}=(\pm \pi/2,\pm \pi/2)$ (CBZ), which are not $C4$-invariant momenta.
However, the wavefunction retains its $s$-wave character~\cite{Bohrdt2021,Bermes2024}; thus to good approximation we can assume that the ground state of the magnetic polaron has no internal excitations and admits a set of ro-vibrational quantum numbers, i.e.~$n=m_4=m_3^{(N)}=0$.
Therefore we assume the low-energy physics of the doped Mott insulator to only contain magnetic polarons in their internal ground state denoted by~$\hat{\pi}^\dagger_{\mathbf{k},\sigma}$; hence we only consider a \textit{single} open channel.

So far, we have solved for the chargon (string) wavefunction~$\ket{\psi_{\rm{c}}}$, which describes a light chargon bound to an infinitely heavy spinon.
However, the spinon can become dispersive via (i) Trugman loops and (ii) spin flip-flop processes.
Taking these processes into account, a dispersion relation~$\varepsilon_\mathrm{sc}(\mathbf{k)}$ for the magnetic polaron can be calculated accurately. For the case of hole doping, this gives rise to the observed hole pockets centered around the nodal points~$\mathbf{k}=(\pm\pi/2,\pm\pi/2)$.

This dispersion relation~$\varepsilon_\mathrm{sc}(\mathbf{k)}$ can be obtained within our truncated basis approach by taking into account Trugman loops and spin flip-flop~$J_\perp$ processes~\cite{Bermes2024}.
Moreover, previous studies have derived the dispersion using several methods, including $1/S$~expansion~\cite{Kane1989,Sachdev1989,Martinez1991} and semi-classical theories~\cite{Shraiman1988a}, which find the following approximate expression:
\begin{align} \label{eq:sc-dispersion}
\begin{split}
    \varepsilon_\mathrm{sc}(\mathbf{k)} &= A [ \cos{(2k_x)} + \cos{(2k_y)} ] \\
    &+ B [ \cos{(k_x + k_y)} + \cos{(k_x - k_y)} ].
\end{split}
\end{align}
Here, the parameters~$A$ and~$B$ are used as fit parameters; from numerical studies of the single hole problem using a self-consistent Born approximation~\cite{Martinez1991} we extract~$A=0.31$ and~$B=0.44$ for realistic cuprate material parameters, $t/J = 10/3$. This corresponds to elliptical hole pockets with mass ratio~$6:1$ at low doping.

In the low-doping regime, the hole pocket-like Fermi surface~\cite{Kane1989,Sachdev1989} has been observed in ARPES studies of hole-doped cuprate compounds~\cite{Kunisada2020,Kurokawa2023} consistent with quantum oscillation measurements~\cite{DoironLeyraud2007}.
As argued in Ref.~\cite{Homeier2023Feshbach}, we assume that at finite but low doping, magnetic polarons can be treated as free fermions forming a Fermi liquid, described by the Hamiltonian
\begin{align} \label{eq:Hopen}
    \H_{\mathrm{open}} = \sum_{\mathbf{k},\sigma} \left[ \varepsilon_\mathrm{sc}(\mathbf{k)} -\mu \right]\pd_{\mathbf{k},\sigma} \p_{\mathbf{k},\sigma}.
\end{align}
Upon increasing the chemical potential~$\mu$ the two hole pockets are filled up. Thus, the free fermion description of the magnetic polarons resembles the normal state of the open channel in the Feshbach hypothesis~\cite{Homeier2023Feshbach}.

%%%%%%%%%%%%%%%%%%%%%%%%%%%%%%%%%%%%%%%%%%%%%%%%%%%%%
\subsection{Interacting magnetic polarons: Open channel}
\label{sec:Open-Channel}
%%%%%%%%%%%%%%%%%%%%%%%%%%%%%%%%%%%%%%%%%%%%%%%%%%%%%
For the Feshbach scattering scenario we need to formally define the scattering channels composed of two magnetic polarons~\mbox{(sc)$^2$} (open channel) and the tightly-bound bosonic (cc) mesons (closed channels).
The low-energy scattering in the open channel is described by two magnetic polarons~$\pd_{\mathbf{k},\sigma}$ in their internal ground state, see Eq.~\eqref{eq:Hopen}.
Our main goal is to calculate their scattering length, characterizing their interaction, in the vicinity of a meson Feshbach resonance~\cite{Homeier2023Feshbach}.
As we discuss later, we only consider low-energy, intra-pocket scattering of two fermions located on the same Fermi surface, i.e. with opposite momenta~$\mathbf{k}_{\uparrow} = -\mathbf{k}_{\downarrow}$, total momentum $\mathbf{Q}=\mathbf{0}$ and well-defined $C4$-angular momentum.
Nevertheless, our formalism -- in principle -- allows us to describe inter-pocket scattering leading to ${\rm (sc)}^2$ pairs with non-zero total momentum~$\mathbf{Q} \neq \mathbf{0}$.

We consider an arbitrary open-channel state with two magnetic polarons, one on each sublattice:
\begin{align} \label{eq:openchannel_planewaves}
    \ket{ \mathbf{k}_\uparrow,\uparrow; \mathbf{k}_\downarrow,\downarrow} = \pd_{\mathbf{k}_\uparrow, \uparrow} \pd_{\mathbf{k}_\downarrow, \downarrow} \ket{0} \in \mathscr{H}_\mathrm{open},
\end{align}
where $\mathscr{H}_\mathrm{open}$ is the open channel Hilbert space.
Using the expression in Eq.~\eqref{eq:sc_boundstates} and assuming no internal ro-vibrational excitations, the pair wavefunction reads
\begin{widetext}
\begin{align} \label{eq:openchannel_wavefunction}
\begin{split}
    \ket{ \mathbf{k}_\uparrow,\uparrow; \mathbf{k}_\downarrow,\downarrow} = \frac{2}{L^2} &\left[ \sum_{\mathbf{j}_\uparrow}e^{-i\mathbf{k}_\uparrow (\mathbf{j}_\uparrow+\mathbf{r})}\sum_{\ell_\uparrow}\sum_{M_{\uparrow}=0}^{\ell_\uparrow-1} \sum_{\lambda_\uparrow^{(0)}}...\sum_{\lambda_\uparrow^{(M_\uparrow)}} \frac{1}{\sqrt{4}}\frac{1}{\sqrt{3^{M_\uparrow}}} \psi_{\rm{sc}}(\mathbf{k}_\uparrow,\ell_\uparrow) \ket{\mathbf{j}_\uparrow, \ell_\uparrow, \{\lambda^{(N)}_\uparrow \} } \right] \\
    \otimes &\left[ \sum_{\mathbf{j}_\downarrow}e^{-i\mathbf{k}_\downarrow \mathbf{j}_\downarrow}\sum_{\ell_\downarrow}\sum_{M_{\downarrow}=0}^{\ell_\downarrow-1} \sum_{\lambda_\downarrow^{(0)}}...\sum_{\lambda_\downarrow^{(M_\downarrow)}} \frac{1}{\sqrt{4}}\frac{1}{\sqrt{3^{M_\downarrow}}} \psi_{\rm{sc}}(\mathbf{k}_\downarrow,\ell_\downarrow) \ket{\mathbf{j}_\downarrow, \ell_\downarrow, \{\lambda^{(N)}_\downarrow \} } \right] ,
\end{split}
\end{align}
\end{widetext}
where we have omitted the ro-vibrational quantum numbers of the amplitudes~$\psi_{\rm{sc}}(\mathbf{k}_\sigma,\ell_\sigma) \equiv \psi_{\rm{sc}}^{(n=0)}(\mathbf{k}_\sigma,\ell_\sigma, m_4=0, {m^{(N)}_3=0})$.
The two spinons, $\mathbf{j}_\uparrow$ and $\mathbf{j}_\downarrow,$ reside on different sublattices that are dislocated by a displacement vector~$\mathbf{r}=(a, 0)$ with lattice spacing~$a$; in the following we set~$a=1$.
In our gauge choice, we assume that the~$\downarrow$-spinons reside on lattice sites~$\mathbf{j}$ and the~$\uparrow$-spinons are displaced by~$\mathbf{r}$, see Fig.~\ref{figMBZ}a (right).

Since we want to describe the low-energy properties of the charge carriers, i.e. magnetic polarons, the relevant scattering predominantly happens at the Fermi surface between a pair of (sc)'s with total (quasi)momentum~\mbox{$\mathbf{Q}\,\mathrm{mod}\mathbf{G}^M=\mathbf{0}$}, where the total (quasi)momentum is only defined up to reciprocal lattice vectors~$\mathbf{G}^M=(\pm\pi/\sqrt{2}, \pm \pi/\sqrt{2})$ in the MBZ.
We expect that the pairs of magnetic polarons we consider in the scattering problem will form Cooper pairs after integrating out the closed channel. 
Therefore, we restrict our following calculations to scattering, or Cooper pairs, in the spin singlet channel.  

Now, we define the zero-momentum singlet pairing field operator, which we expand in angular momentum eigenfunctions
\begin{align} 
    \hat{\Delta}^\dagger_{m_4} (\mathbf{Q}=\mathbf{0}) &= \frac{1}{\sqrt{2}}\sum_{\mathbf{k}} f_{m_4}(\mathbf{k})\left( \pd_{-\mathbf{k},\uparrow}\pd_{\mathbf{k},\downarrow} - \pd_{-\mathbf{k},\downarrow}\pd_{\mathbf{k},\uparrow}  \right)
\end{align}
with the relative momentum~$\mathbf{k}$, which creates two magnetic polarons in the open channel from vacuum.
The function~$f_{m_4}(\mathbf{k})$ is an eigenfunction of the $C4$-rotation operator, i.e. it transforms as $f_{m_4}(\mathbf{k}) \rightarrow e^{-i m_4 \pi/2}f_{m_4}(\mathbf{k})$ under $\pi/2$~rotations; otherwise the exact functional form of~$f_{m_4}(\mathbf{k})$ is arbitrary.
We note that the angular momentum~$m_4$ refers to the \emph{orbital} angular momentum of the pair, not to the magnetic polaron's internal degrees-of-freedom.
By using the fermionic anticommutation relations of the magnetic polarons~$\pd_{\mathbf{k},\sigma}$, we find
\begin{align} \label{eq:fermionic-openchannel}
    \hat{\Delta}^\dagger_{m_4} (\mathbf{Q}=\mathbf{0}) = \frac{1}{\sqrt{2}}\sum_{\mathbf{k}} (1 +  e^{-i m_4 \pi})f_{m_4}(\mathbf{k})\pd_{-\mathbf{k},\uparrow}\pd_{\mathbf{k},\downarrow}, 
\end{align}
which is only non-zero for even parity $s$-wave ($m_4=0$) and \mbox{$d$-wave} ($m_4=2$) open channel states.
In summary, the low-energy scattering of spin singlet and zero momentum pairs restricts the open channel states to have $s$-wave or $d$-wave spatial angular momentum.

%%%%%%%%%%%%%%%%%%%%%%%%%%%%%%%%%%%%%%%%%%%%%%%%%%%%%
\subsection{Chargon-chargon (cc) bound state:\\*Closed channel}
\label{sec:cc-stringtheory}
%%%%%%%%%%%%%%%%%%%%%%%%%%%%%%%%%%%%%%%%%%%%%%%%%%%%%
Next, we discuss the properties, and evidence for, tightly-bound bosonic mesons formed by chargon-chargon (cc) bound states.
These are the constituents of the closed channel in the Feshbach model proposed in Ref.~\cite{Homeier2023Feshbach}.
We closely follow the derivation in Ref.~\cite{Grusdt2023}.

As for the (sc) bound states, the light chargons displace the frozen spin background due to their fast motion~$\propto t$, see Fig.~\ref{figBethe}a (bottom).
However, one chargon can retrace the path of the other giving rise to bound states.
Because the chargons are indistinguishable, spinless fermions, the particle statistics plays a crucial role\footnote{Alternatively, one may treat chargons as bosons, but in this case the fermionic statistics of the underlying spins in the Hubbard or \mbox{$t$-$J$}~model lead to an additional statistical phase associated with the geometric string of displaced spins connecting the two chargons. This ultimately leads to an equivalent description~\cite{Grusdt2023}. The chargon-chargon bound state is of bosonic nature.}.

Again, we apply the frozen spin approximation, i.e. we consider two holes or doublons created on opposite sublattices in a N\'eel ordered state, and consider their correlated motion through the background.
In particular, we first assume two distinguishable chargons labeled~$A$ and $B$, and perform an antisymmetrization procedure afterwards.

In contrast to the (sc) bound states, where the spinon was considered heavy and thus frozen, now both constituents~$A$ and~$B$ are mobile.
Hence, we perform a Lee-Low-Pines transformation~\cite{Lee1953} into the co-moving frame of chargon~$A$, which yields a model of a single chargon at one end of a string, in an effective string potential and with an effective tunneling amplitude that depends on the total momentum~$\mathbf{Q}$.

Analogously to the (sc) case, we can introduce a set of orthonormal basis states~$\ket{\mathbf{x}_{\mathrm{c}}, \Sigma_{\mathrm{cc}}}$ with the position~$\mathbf{x}_{\mathrm{c}}$ of chargon~$A$ and the string~$\Sigma_{\mathrm{cc}}$ connecting the two chargons.
Again, these states are defined on a Bethe lattice, see Fig.~\ref{figBethe}b, but with chargon~$A$ in the center.
Thus, we can likewise use basis states with fixed string length~$\ell$ and angles~$\{ \lambda^{(N)} \}$ as above,
\begin{align}
    \ket{\mathbf{x}_{\mathrm{c}}, \Sigma_{\mathrm{cc}}} = \ket{\mathbf{x}_{\mathrm{c}}, \ell, \{ \lambda^{(N)} \} }
\end{align}
and perform the Fourier transformation to obtain momentum states
\begin{align}
\begin{split}
    \ket{\mathbf{Q}, \ell,  \{ \lambda^{(N)} \}} = \frac{1}{\sqrt{L^2}}\sum_{\mathbf{x}_{\mathrm{c}}} e^{-i \mathbf{Q} \mathbf{x}_{\mathrm{c}}} \ket{\mathbf{x}_{\mathrm{c}}, \ell, \{ \lambda^{(N)} \}}.
\end{split}
\end{align}
Note that the momentum~$\mathbf{Q}$ is now defined in the CBZ because, at the level of our approximations so far, the chargons are not restricted to a sublattice.
Further, at $C4$-invariant momenta, we can define rotational eigenstates
\begin{align} \label{eq:basis_m4m3_cc}
\begin{split}
     &\ket{\mathbf{Q}, \ell, m_4, \{ m_3^{(N)} \}} = \\
     &=\sum_{M = 0}^{\ell-1} \frac{1}{\sqrt{4}}\frac{1}{\sqrt{3^M}} e^{-i\left[ \lambda^{(0)}m_4 + \sum_{N=1}^M \lambda^{(N)} m_3^{(N)} \right]} \ket{\mathbf{Q}, \ell,  \{ \lambda^{(N)} \}}.
\end{split}
\end{align}
To accommodate for the particle statistics, the states in Eq.~\eqref{eq:basis_m4m3_cc} have to be (anti)symmetrized.
The resulting antisymmetric states then span the closed channel Hilbert space~$\mathscr{H}_\mathrm{closed}$.

At the two $C4$-invariant momenta, $\mathbf{Q}=\mathbf{0}$ and $\mathbf{Q}=\bm{\pi}$, the chargon-chargon bound states have well-defined rotational quantum numbers with $p$- and \mbox{$f$-wave} ($\mathbf{Q}=\mathbf{0}$) as well as $s$- and \mbox{$d$-wave} ($\mathbf{Q}=\bm{\pi}$) symmetry in the fermionic sector.

The translational and $C4$-rotational invariance of the underlying model, Eq.~\eqref{eq:tJ}, allow us to derive selection rules for the matrix elements, which couple between open and closed channel states.
As discussed above, we only consider couplings to states with $\mathbf{Q}\,\mathrm{mod}\mathbf{G}^\mathrm{M}=\mathbf{0}$; thus the channels have well-defined angular momentum quantum numbers.
Since the open channel has even parity, Eq.~\eqref{eq:fermionic-openchannel}, we conclude that only even parity channels contribute to the scattering of magnetic polarons; hence non-zero matrix elements arise only between $s$-wave (\mbox{$d$-wave}) open channel and  $s$-wave (\mbox{$d$-wave}) closed channel states at~$\mathbf{Q}\,\mathrm{mod}\mathbf{G}^\mathrm{M}=\bm{\pi}\,\mathrm{mod}\mathbf{G}^\mathrm{M}=\mathbf{0}$.

\begin{figure*}[t!]
\includegraphics[width=0.95\linewidth]{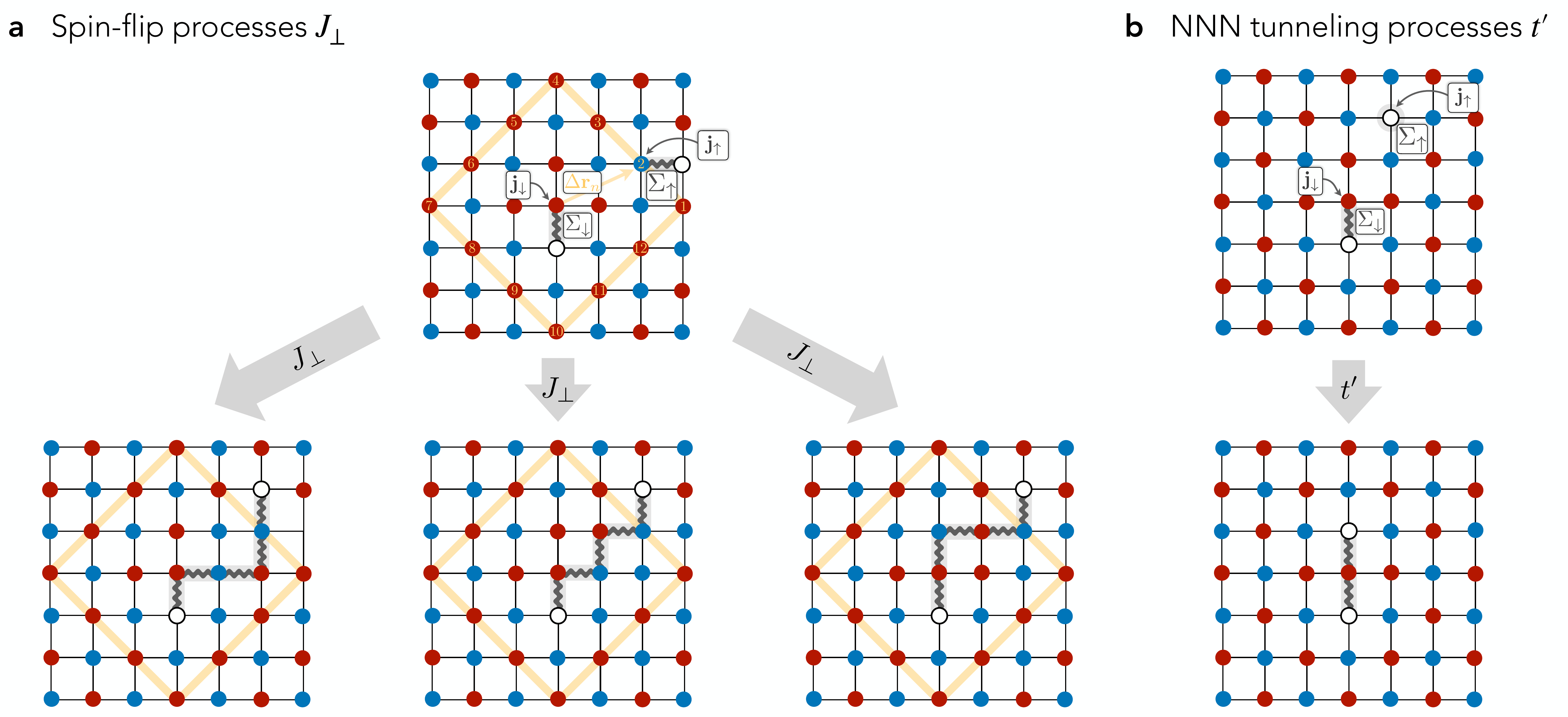}
\caption{\textbf{Recombination processes: sc $+$ sc $\rightarrow$ cc.} The open and closed channel are coupled via processes that annihilate and create pairwise spinons. \textbf{a} Spin-flip processes couple magnetic polarons at Manhattan distance~$||\Delta \mathbf{r}_{n}||_M =||\mathbf{j}_\uparrow -(\mathbf{j}_\downarrow+\mathbf{r}) ||_M=3$. In the co-moving frame of one spinon, the other spinon has to be located on sites indicated by the yellow box. The spin flip-flop~$J_\perp$ annihilates the magnetic polarons and creates a chargon-chargon pair with string length~$\ell = \ell_\downarrow + \ell_\uparrow+3$ (besides one special case with $\ell = \ell_\downarrow + \ell_\uparrow-1$). For the shortest string length approximation (SSLA) we consider~$\ell_\downarrow = \ell_\uparrow = 0$. \textbf{b} Similarly, next-nearest neighbour tunneling~$t'$ couples the open and closed channel.  }
\label{figRecombination}
\end{figure*}

A Feshbach resonance describes the scattering within an open channel in the presence of a near-resonant closed channel that can be virtually occupied.
In addition to the coupling matrix elements, the bare energy difference~$\Delta E_{m_4}$ between the (uncoupled) channels determines the strength~$\propto 1/\Delta E_{m_4}$ of the scattering length.
Here, we distinguish the two possible couplings to closed channels with $s$-wave ($m_4=0$) and $d$-wave ($m_4=2$) symmetry; hence we further reduce the multichannel description using selection rules.
Numerical simulations of the $t$-$J$ model~\cite{Poilblanc1994, Bohrdt2023_dichotomy} have calculated the angular-momentum resolved two-hole spectra at $\mathbf{Q}\,\mathrm{mod}\mathbf{G}^\mathrm{M}=\mathbf{0}$.
At the relevant momenta, they clearly indicate a large energy difference between the $s$-wave, $\Delta E_{m_4=0}=\mathcal{O}(t)$, and $d$-wave, $\Delta E_{m_4=2}=\mathcal{O}(J)$, channels; hence only the $d$-wave (cc) state gives rise to near-resonant scattering.
Therefore, we conclude that the effective scattering length is dominated by the $d$-wave channel, which allows us to consider an effective two-channel model in the following~\cite{Homeier2023Feshbach}.
Likewise, we recognize that other scenarios are possible such as triplet pairing or inter-valley scattering, which could be described analogously within our formalism.

Now, we define the basis states~$\ket{\mathbf{Q}=\bm{\pi}, \ell, m_4=2}$ to be the fermionic states in the relevant two-channel model.
This allows us to express the closed-channel wavefunction of the tightly-bound (cc) state without vibrational excitations,
\begin{align} \label{eq:cc_realspace}
\begin{split}
    &\ket{\mathbf{Q}=\bm{\pi}} = \hat{b}^\dagger_{m_4=2}(\mathbf{Q}=\bm{\pi})\ket{0} = \\
    &= \frac{1}{\sqrt{L^2}} \sum_{\mathbf{x}_{\mathrm{c}}} e^{-i \mathbf{Q} \mathbf{x}_{\mathrm{c}}} \sum_{\ell} \phi_{\rm{cc}}(\mathbf{Q}=\bm{\pi}, \ell)\ket{\mathbf{Q}=\bm{\pi}, \ell, m_4=2}.
\end{split}
\end{align}
Here, we have defined the bosonic creation operator for the (cc) state~$\hat{b}^\dagger_{m_4=2}(\mathbf{Q})$ with angular momentum~$m_4=2$ and total momentum~$\mathbf{Q}=\bm{\pi}$.
Further, we can expand the wavefunction in the angular basis of string states, Eq.~\eqref{eq:basis_m4m3_cc},
\begin{align} \label{eq:cc_realspace-angle}
\begin{split}
    &\ket{\mathbf{Q}=\bm{\pi}} = \\
    &= \frac{1}{\sqrt{L^2}} \sum_{\mathbf{x}_{\mathrm{c}}} e^{-i \mathbf{Q} \mathbf{x}_{\mathrm{c}}} \sum_{\ell} \sum_{M=0}^{\ell-1}\sum_{\lambda^{(0)}}...\sum_{\lambda^{(M)}} e^{-i \lambda^{(0)} m_4} \\
    &\times \phi_{\rm{cc}}(\mathbf{Q}=\bm{\pi}, \ell)\hat{P}_f\ket{\mathbf{x}_{\mathrm{c}}, \ell, \{ \lambda^{(N)} \} } ,
\end{split}
\end{align}
where~$\hat{P}_f$ projects onto the fermionic states.

In the next section, we will describe the coupling between the open and closed channels.
To this end, we need to describe the open and closed channel on an equal footing.
In the geometric string picture, the (cc) bound state does not distinguish between the sublattices and therefore, in this formulation, the momenta can be defined in the CBZ; in contrast to the magnetic polarons, which have a well-defined sublattice/spin quantum number.
Therefore, we fold the momentum from the larger CBZ into the smaller MBZ, see Fig.~\ref{figMBZ}b, by introducing a band index~$\alpha=0,1$, such that
\begin{align}
    \hat{b}^\dagger_{m_4=2,\alpha}(\mathbf{Q}) = \sqrt{\frac{2}{L^2}}\sum_{\mathbf{j}_\downarrow}e^{-i (\mathbf{j}_\downarrow+\alpha \mathbf{r}) \mathbf{Q}} \hat{b}^\dagger_{m_4=2,\alpha}(\mathbf{j}_\downarrow),
\end{align}
where~$\mathbf{j}_\downarrow$ now labels the two site unit cells and~$\alpha$ is a band index describing the position within the unit cell.
In particular, we can consider the (cc) creation operator in the larger CBZ and relate it to the MBZ by considering
\begin{widetext}
\begin{align} \label{eq:sinlget-triplet-closedchannel}
\begin{split}
    \hat{b}^\dagger_{m_4=2}(\mathbf{Q}) &= \frac{1}{\sqrt{L^2}} \sum_{\mathbf{j}_\downarrow} e^{-i \mathbf{j}_\downarrow \mathbf{Q}} \left[ \hat{b}^\dagger_{m_4=2,0}(\mathbf{j}_\downarrow) + e^{-i\mathbf{r}\mathbf{Q}}\hat{b}^\dagger_{m_4=2,1}(\mathbf{j}_\downarrow) \right] = \\
    &= \begin{cases}  \frac{1}{\sqrt{2}}\big[ \hat{b}^\dagger_{m_4=2,0}(\mathbf{Q}) + \hat{b}^\dagger_{m_4=2,1}(\mathbf{Q}) \big]  &\text{for} ~ \mathbf{Q} \in \mathrm{MBZ} \\ &\\\frac{1}{\sqrt{2}} \big[ \hat{b}^\dagger_{m_4=2,0}(\mathbf{Q}\,\mathrm{mod}\mathbf{G}^\mathrm{M}) + e^{i \mathbf{r} \mathbf{G}^\mathrm{M}}\hat{b}^\dagger_{m_4=2,1}(\mathbf{Q}\,\mathrm{mod}\mathbf{G}^\mathrm{M}) \big]  ~~ &\text{for} ~ \mathbf{Q} \notin \mathrm{MBZ} \end{cases}
\end{split}
\end{align}
\end{widetext}
Therefore, we find that the momentum $\mathbf{Q}=\mathbf{0}$ ($\mathbf{Q}=\bm{\pi}$) corresponds to triplet (singlet) combinations in the band index sector; hence the band index equips the (cc) bound state with a pseudospin.
This ultimately allows us to couple an open channel pair, see Eq.~\eqref{eq:fermionic-openchannel}, with $s$- and \mbox{$d$-wave} angular momentum to a closed channel state despite their constituents being spinful (open channel) and spinless (closed channel).

Note that throughout Sec.~\ref{sec:OpenClosed}, we considered a classical N{\'e}el background, i.e. $J_\perp =0$, to derive the open and closed channel states. However the two channels still exist in the presence of spin fluctuations~$J_\perp$ and only microscopic details are affected~\cite{Kane1989, Beran1996, Grusdt2019, Bohrdt2023_dichotomy}.
Further, cold atom experiments in the Fermi-Hubbard model have shown signatures of strings~\cite{Chiu2019,Koepsell2019,Bohrdt2019} indicating that the geometric picture is valid beyond the $t$-$J$ model and at finite temperature.
While in the following we will assume a perfect product N{\'e}el state~$\ket{0}$, we emphasize that strong local AFM correlations with coherence length of~$\xi_{\rm{AFM}}/a \gtrsim 10$ should lead to qualitatively similar result.

%%%%%%%%%%%%%%%%%%%%%%%%%%%%%%%%%%%%%%%%%%%%%%%%%%%%%
\section{Meson scattering interaction}
\label{sec:ScattLength}
%%%%%%%%%%%%%%%%%%%%%%%%%%%%%%%%%%%%%%%%%%%%%%%%%%%%%

In the proposed Feshbach scenario~\cite{Homeier2023Feshbach}, it is suggested that a magnetic polaron pair~$\rm{(sc)}^2$ and a (cc) meson can spatially overlap leading to a coupling of the two channels.
Here, we explicitly calculate the coupling matrix elements (or form factors) originating from the microscopic Hamiltonian~\eqref{eq:tJ} by applying the open and closed channel string description introduced in Secs~\ref{sec:Open-Channel} and~\ref{sec:cc-stringtheory}. 
The possible coupling processes between the two channels are associated with 1) spin-flip processes~($J_\perp$) and 2) NNN tunneling~($t'$), as illustrated in Fig.~\ref{figRecombination} and defined by the open-closed channel coupling Hamiltonian
\begin{align}
\begin{split} \label{eq:Ham-openclosed}
    \H_{\rm{oc}} &= \H_{J_\perp} + \H_{t'} \\
    \H_{J_\perp} &=\frac{J_\perp}{2} \sum_{\ij} \left( \hat{S}_i^+ \hat{S}_j^- + \mathrm{H.c.} \right) \\
    \H_{t'} &= -t' \sum_{\langle\!\ij\!\rangle} \sum_{\sigma} \left( \cd_{\mathbf{i},\sigma}\c_{\mathbf{j},\sigma} +\mathrm{H.c.}  \right),
\end{split}
\end{align}
projected to our truncated two-channel basis.
The resulting scattering interaction~$V_{\mathbf{k},\mathbf{k}'}$ describing the Feshbach resonance is then given by
\begin{align} \label{eq:Feshbach}
    V_{\mathbf{k},\mathbf{k}'} &= \frac{1}{L^2}\sum_{m_4}\frac{\mathcal{M}_{m_4}^*(\mathbf{k'})\mathcal{M}_{m_4}(\mathbf{k})}{\Delta E_{m_4}}
\end{align}
with the form factors/matrix elements~$\mathcal{M}_{m_4}(\mathbf{k})$.
In the following, we focus on the relevant \mbox{$d$-wave} scattering channel ($m_4=2$), i.e. we want to evaluate
\begin{subequations} \label{eq:scatteringlength_general}
\begin{align} 
    \mathcal{M}_2(\mathbf{k}) &= J_\perp \mathcal{M}_2^{J_\perp}(\mathbf{k}) + t'\mathcal{M}_2^{t'}(\mathbf{k}) \\
    \frac{\kappa}{\sqrt{L^2}}\mathcal{M}_2^{\kappa}(\mathbf{k}) &= \bra{0}\hat{b}_{m_4=2}(\mathbf{Q}=\bm{\pi})\H_{\kappa} \pd_{-\mathbf{k},\uparrow}\pd_{\mathbf{k},\downarrow} \ket{0}, \label{eq:matrixelement_curlyM}
\end{align}
\end{subequations}
where~$\kappa = J_\perp, t'$ and $\mathbf{k}'$ ($\mathbf{k}$) is the in-coming (out-going) momentum.
From Eq.~\eqref{eq:scatteringlength_general}, we find that it is sufficient to calculate the matrix elements for spin-flip and NNN tunneling processes individually.

%%%%%%%%%%
% begin Table
%%%%%%%%%%
{\renewcommand{\arraystretch}{2}
\begin{table*}\centering
\begin{tabularx}{\textwidth} { | >{\centering\arraybackslash}X  | >{\centering\arraybackslash}X  | >{\centering\arraybackslash}X  | >{\centering\arraybackslash}X  | >{\centering\arraybackslash}X  | >{\centering\arraybackslash}X  | >{\centering\arraybackslash}X  | }
 \hline
  & $\ell=0$ & $\ell=1$ & $\ell=2$ & $\ell=3$ & $\ell=4$ & $\ell=5$ \\
 \specialrule{.2em}{.1em}{.1em} 
 $\psi_{\rm{sc}}(\mathbf{k}, \ell)$ & $\sqrt{0.25}$  & $\sqrt{0.38/4}$  & $\sqrt{0.22/(4\cdot3)}$ & $\sqrt{0.09/(4\cdot3^2)}$ & $\sqrt{0.05/(4\cdot3^3)}$ & $\sqrt{0.01/(4\cdot3^4)}$ \\
 \hline
  $p_{\mathbf{k}, \rm{sc}}(\ell)$ & $0.25$  & $0.38$  & $0.22$ & $0.09$ & $0.05$ & $0.01$ \\
  \specialrule{.2em}{.1em}{.1em} 
  $\phi_{\rm{cc}}(\mathbf{Q}=\bm{\pi},\ell)$ & -- & $\sqrt{0.09/4}$ & $\sqrt{0.26/8}$ & $\sqrt{0.32/20}$ & $\sqrt{0.22/48}$ & $\sqrt{0.09/148}$ \\
  \hline
  $p_{\mathbf{Q}=\bm{\pi},\rm{cc}}(\ell)$ & -- & $0.09$ & $0.26$ & $0.32$ & $0.22$ & $0.09$ \\
\hline
\end{tabularx}
\caption{\textbf{String length amplitudes.} The wavefunction amplitudes for the spinon-chargon~$\psi_{\rm{sc}}$ are extracted from Ref.~\cite{Grusdt2019} for $t/J=3$. The wavefunction amplitudes for the chargon-chargon~$\phi_{\rm{cc}}$ are extracted from Ref.~\cite{Grusdt2023} for $t/J=3$. The renormalization factor for the latter includes the normalization after projection onto fermionic states at~$\mathbf{Q}=\bm{\pi}$. The string length distribution can be obtained from $p_{\rm{sc}}(\mathbf{k},\ell) = \mathcal{N}^\ell_{\rm{sc}} |\psi_{\rm{sc}}(\mathbf{k},\ell)|^2$, where $\mathcal{N}^\ell_{\rm{sc}}$ are the number of string states of length~$\ell$, and further around the dispersion minimum the momentum dependency is negligible $p_{\rm{sc}}(\mathbf{k},\ell) \approx p_{\rm{sc}}(\ell)$. The string length distribution for the (cc) case is obtained analogously. }
\label{table:string_amplitudes}
\end{table*}
%%%%%%%%%%
% end Table
%%%%%%%%%%

\begin{figure*}[t!]
\includegraphics[width=0.8\linewidth]{figSSLA_reshape-Jperp.pdf}
\caption{\textbf{Shortest string length approximation (SSLA) -- Spin flip.} \textbf{a} We show the contributions to the wavefunction of the open channel, where the two chargons are Manhattan distance~$||\Delta \mathbf{r}_n||_M=3$ apart and the string length is~$\ell_\uparrow=\ell_\downarrow=0$. These states couple to the closed channel with substantial overlap to the (cc) wavefunction illustrated in \textbf{b}, where the string length is~$\ell=3$. The overlaps have to be weighted by momentum-dependent phase factors (here with momenta~$k_x^M,k_y^M$ measured in the $\pi/4$-rotated MBZ basis, see Eq.~\eqref{eq:unitvecs_MBZ}) and symmetry properties of the \mbox{$d$-wave} closed channel have to be taken into account. The shown contributions constitute all twelve terms in the SSLA. }
\label{figSSLA-Jperp}
\end{figure*}
%%%%%%%%%%%%%%%%%%%%%%%%%%%%%%%%%%%%%%%%%%%%%%%%%%%%%
\subsection{Spin-flip processes}
\label{sec:spinflip}
%%%%%%%%%%%%%%%%%%%%%%%%%%%%%%%%%%%%%%%%%%%%%%%%%%%%%
First, we focus on the spin-flip recombination processes, i.e. we calculate the form factor~$\frac{J_\perp}{\sqrt{L^2}}\mathcal{M}^{J_\perp}_2(\mathbf{k})$.
To evaluate Eq.~\eqref{eq:matrixelement_curlyM}, we expand the states in real space according to Eqs.~\eqref{eq:openchannel_wavefunction} and~\eqref{eq:cc_realspace-angle}.
In real space, it is straightforward to determine how the spin flip-flop interactions~$J_\perp$ couple between the open and closed channel states, see Fig.~\ref{figRecombination}a.
To be precise, they annihilate opposite spinons at sites~$\mathbf{j}_\downarrow$ and~$\mathbf{j}_\uparrow$ that are Manhattan distance~$||\Delta \mathbf{r}_{n}||_M =||(\mathbf{j}_\uparrow + \mathbf{r})-\mathbf{j}_\downarrow ||_M=3$ apart.
The recombination processes thus couple magnetic polarons of length~$\ell_\downarrow$ and $\ell_\uparrow$ to (cc) states of length~$\ell=\ell_\downarrow+\ell_\uparrow+\Delta\ell$ with~$\Delta\ell=-1,3$.

For realistic parameters~$t/J\approx 2, 3$, the magnetic polaron~(sc) string length is peaked around~$\ell_\sigma = 0$ and the (cc) string length distribution has its maximum at~$\ell=3$, see Table~\ref{table:string_amplitudes}.
Therefore, the largest contribution to the form factors~\eqref{eq:matrixelement_curlyM} occurs for the peaked string lengths, which justifies a short string length approximation (SSLA). The latter includes only~$\ell_{\uparrow}=\ell_{\downarrow} = 0$ and $\ell = 3$.
Later, we will systematically include longer strings based on numerical calculations of the matrix elements.
In SSLA the matrix elements can be written as
\begin{widetext}
\begin{align}
\begin{split}
   \frac{J_\perp}{\sqrt{L^2}}\mathcal{M}_2^{J_\perp}(\mathbf{k})= 
   \frac{2}{L^{3}} \sum_{\mathbf{x}_{\mathrm{c}}, \mathbf{j}_\downarrow, \mathbf{j}_\uparrow} e^{i \left[ \mathbf{Q} \mathbf{x}_{\mathrm{c}} - \mathbf{k} (\mathbf{j}_\uparrow+\mathbf{r}) + \mathbf{k} \mathbf{j}_\downarrow \right]} &\sum_{\lambda^{(0)}}\sum_{\lambda^{(1)}}\sum_{\lambda^{(2)}}e^{i \lambda^{(0)}m_4 }\phi_{\rm{cc}}^*(\mathbf{Q}, \ell=3) \psi_{\rm{sc}}(\mathbf{k}, \ell_\downarrow=0)\psi_{\rm{sc}}(-\mathbf{k}, \ell_\uparrow=0) \times \\
   &\times\bra{\mathbf{x}_{\mathrm{c}}, \ell=3, \{ \lambda^{(N)}\}}\hat{P}_f \H_{J_\perp} \Big( \ket{\mathbf{j}_\downarrow, \ell_\downarrow=0 } \otimes \ket{\mathbf{j}_\uparrow, \ell_\uparrow=0 } \Big).
\end{split}
\end{align}
In real space the coupling elements are given by
\begin{align}
\begin{split}
    &\bra{\mathbf{x}_{\mathrm{c}}, \ell=3, \{ \lambda^{(N)}\}}\hat{P}_f \H_{J_\perp} \Big( \ket{\mathbf{j}_\downarrow, \ell_\downarrow=0 } \otimes \ket{\mathbf{j}_\uparrow, \ell_\uparrow=0 } \Big) =J_\perp \delta_{\mathbf{x}_{\mathrm{c}},\mathbf{j}_\downarrow} \sum_{\Delta \mathbf{r}_{n}} \delta_{\Delta \mathbf{r}_n, \mathbf{j}_\uparrow+\mathbf{r}-\mathbf{j}_\downarrow} \sum_{\{\lambda{(N)}\}}\delta_{\lambda^{(N)},\lambda^{(N)}_f},
\end{split}
\end{align}
where the last term ensures that only fermionic string states~$\lambda^{(N)}_f$ contribute and $\Delta \mathbf{r}_n$ denotes all real space configurations that can annihilate spinons, see Fig.~\ref{figRecombination}a.
The non-zero matrix elements can be read off from the open and closed channel wavefunctions illustrated in Fig.~\ref{figSSLA-Jperp}.

Therefore, the expression becomes
\begin{align} \label{eq:formfactor-long}
\begin{split}
    \frac{J_\perp}{\sqrt{L^2}}\mathcal{M}_2^{J_\perp}(\mathbf{k})= \frac{2 J_\perp}{L^{3}} \sum_{\mathbf{j}_\downarrow} e^{ i \mathbf{j}_\downarrow \mathbf{Q} } \phi_{\rm{cc}}^*(\mathbf{Q}, \ell=3) \psi_{\rm{sc}}(\mathbf{k}, \ell_\downarrow=0)\psi_{\rm{sc}}(-\mathbf{k}, \ell_\uparrow=0)\underbrace{\sum_{\Delta \mathbf{r}_n} e^{-i \mathbf{k} \Delta \mathbf{r}_n} \sum_{\lambda^{(0)}}{}^{'} \sum_{\lambda^{(1)}}{}^{'} \sum_{\lambda^{(2)}}{}^{'}  e^{i \lambda^{(0)}m_4 }}_{\equiv\chi^{J_\perp}(\mathbf{k})}.
\end{split}
\end{align}
\end{widetext}
The restricted sum runs over the fermionic string states with strings starting at~$\mathbf{j}_\downarrow$ and ending at~$\mathbf{j}_\downarrow+\Delta \mathbf{r}_n$.
To simplify the expression, we use the following identity
\begin{align}
    \sum_{\mathbf{j}_\downarrow} e^{- i \mathbf{Q} \mathbf{j}_\downarrow} = \frac{L^2}{2}\delta \left[ \mathbf{Q}\,\mathrm{mod}\mathbf{G}^\mathrm{M}\right],
\end{align}
which gives the expression for the form factors.

We further define the functions
\begin{align}
    \Omega(\mathbf{Q}, \mathbf{k}) &= \phi_{\rm{cc}}^*(\mathbf{Q}, \ell=3) \psi_{\rm{sc}}(\mathbf{k}, \ell_\downarrow=0)\psi_{\rm{sc}}(-\mathbf{k}, \ell_\uparrow=0) \label{eq:Omega} \\
    \chi^{J_\perp}(\mathbf{k}) &= \sum_{\Delta \mathbf{r}_n} e^{-i \mathbf{k} \Delta \mathbf{r}_n} \sum_{\lambda^{(0)}}{}^{'} \sum_{\lambda^{(1)}}{}^{'} \sum_{\lambda^{(2)}}{}^{'}  e^{i 2 \lambda^{(0)} },
\end{align}
where we have set~$m_4=2$ for the \mbox{$d$-wave} channel.
In the vicinity of the dispersion minimum,~$\mathbf{k}=(\pm \pi/2, \pm\pi/2) + \delta \mathbf{k}$ with~$|\delta \mathbf{k}| \ll \pi$, the wavefunction amplitudes are assumed to be $\mathbf{k}$-independent for $s$-wave magnetic polarons; in Sec.~\ref{sec:truncated-basis} we account for the full momentum dependence.

\begin{figure*}[t!]
\includegraphics[width=1\linewidth]{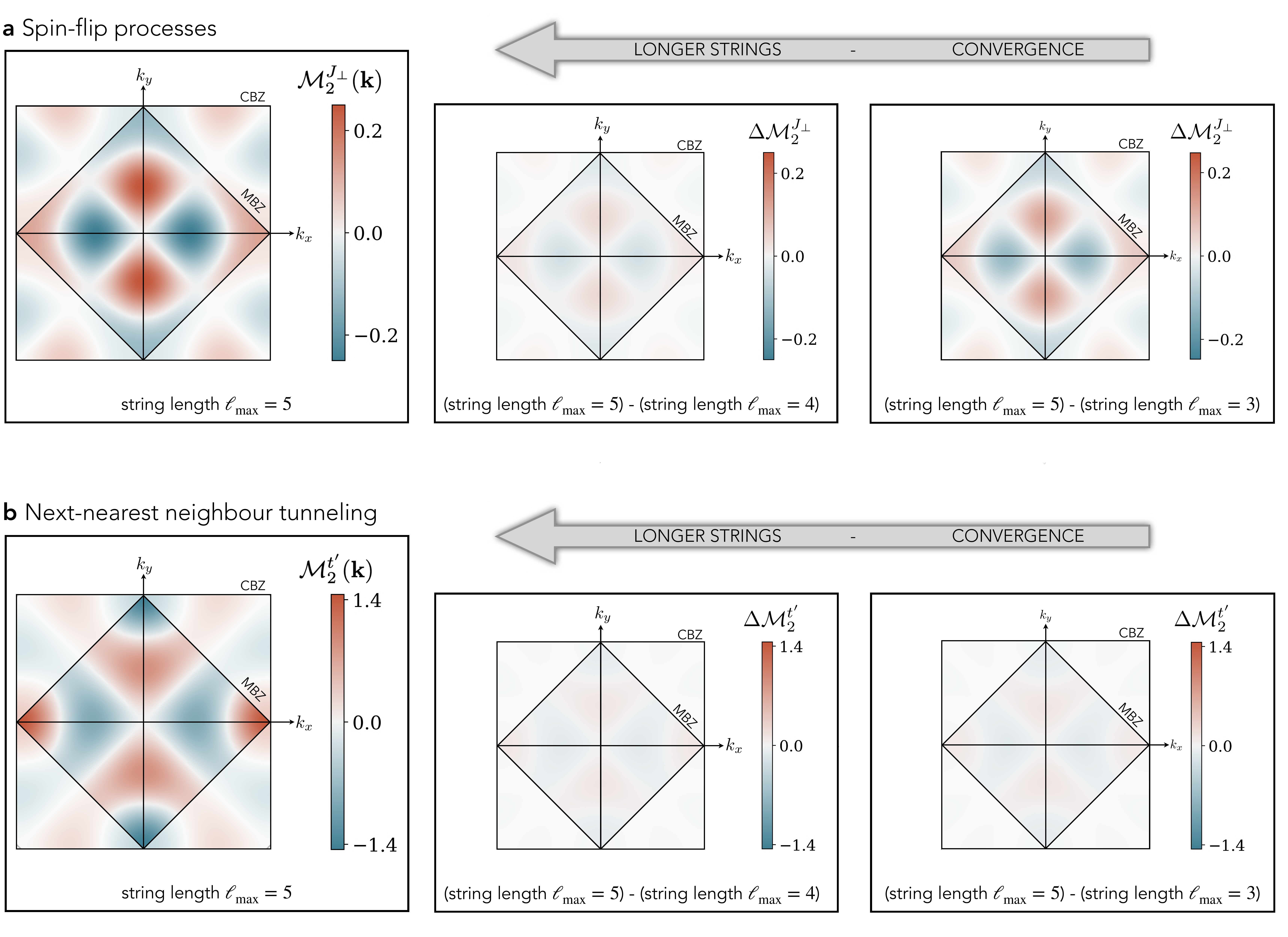}
\caption{\textbf{Testing the shortest string length approximation (SSLA).} We compute the matrix element relevant for the scattering processes numerically, and cut off at different maximal string length~$\ell_{\rm{max}}$ of the involved (cc) states for \textbf{a} spin-flip processes and \textbf{b} NNN tunneling. We compare the $\ell_\mathrm{max}=5$ calculations to calculations with shorter strings. We note that the relative difference between $\ell_\mathrm{max}=5$ and $\ell_\mathrm{max}=3$ is relatively small, which justifies the analytically tractable SSLA approximation. }
\label{figSSLA_approx}
\end{figure*}

Instead, the function~$\chi^{J_\perp}(\mathbf{k})$ is highly $\mathbf{k}$-dependent and determines the structure of the form factor Eq.~\eqref{eq:scatteringlength_general}.
In particular, in SSLA we can evaluate the form factors for spin-flip recombination processes analytically.
We carefully treat the momentum~$\mathbf{k}^M$ and real space vectors~$\Delta \mathbf{r}_n$ in the MBZ, see Fig.~\ref{figMBZ} and Eq.~\eqref{eq:unitvecs_MBZ}.
In Fig.~\ref{figSSLA-Jperp}, we show the corresponding phase factors~$\chi^{J_\perp}(\mathbf{k})$ for momenta in the MBZ.
We sum up the matrix element, and obtain
\begin{align}
\begin{split}
    \chi^{J_\perp}(\mathbf{k}^M) = &2\Bigg[ \cos{\left(\frac{k_x^M-3k_y^M}{\sqrt{2}}\right)} + \cos{\left(\frac{3k_x^M-3k_y^M}{\sqrt{2}}\right)} \\
    &+ \cos{\left(\frac{3k_x^M-k_y^M}{\sqrt{2}}\right)} 
    - \cos{\left(\frac{3k_x^M+3k_y^M}{\sqrt{2}}\right)}\\
    &- \cos{\left(\frac{3k_x^M+k_y^M}{\sqrt{2}}\right)} - \cos{\left(\frac{k_x^M+3k_y^M}{\sqrt{2}}\right)}\Bigg],
\end{split}
\end{align}
which has $\rm{d}_{x^2-y^2}$ nodal structure in the CBZ (or equivalently $\rm{d}_{xy}$ nodal structure in the $\pi/4$-rotated MBZ).

It is important to confirm the validity of the SSLA by systematically including longer strings. 
We automatize the formalism described above and perform exact numerical calculations in the string picture.
Each string length realization now has to be weighted by wavefunction amplitudes~$\phi_{\rm{cc}}(\mathbf{Q}=\bm{\pi}, \ell)$ and $\psi_{\rm{sc}}(\mathbf{k}, \ell_\sigma)$, which we extract from previous studies, see Refs.~\cite{Grusdt2019, Grusdt2023} and Table~\ref{table:string_amplitudes}.

We calculate the matrix element and plot~$\mathcal{M}_2^{J_\perp}(\mathbf{k})$, see Eq.~\eqref{eq:matrixelement_curlyM}, for different maximal (cc) string length cut-offs~$\ell_\mathrm{max}$ up to~$\ell_\mathrm{max}=5$.
Then, we compare the calculation of~$\ell_\mathrm{max}=5$ with short strings as shown in Figure~\ref{figSSLA_approx}a.
Note that for $\ell_\mathrm{max}=5$, the string lengths of the magnetic polarons are bounded to~$\mathrm{max}(\ell_\downarrow,\ell_\uparrow)=2$, which has the advantage that we do not have to consider Trugman loops or crossings of strings.

We find that already the SSLA gives qualitatively the correct behaviour, while the quantitative results are only slightly renormalized by including longer string lengths.
We emphasize the robustness of the proposed Feshbach mechanism~\cite{Homeier2023Feshbach} with respect to the $d_{x^2-y^2}$ nodal structure, which is caused by the symmetry properties of the closed channel but not by the details of the geometric string wavefunctions.

%%%%%%%%%%%%%%%%%%%%%%%%%%%%%%%%%%%%%%%%%%%%%%%%%%%%%
\subsection{Next-nearest neighbour tunnelings}
\label{sec:NNNtunneling}
%%%%%%%%%%%%%%%%%%%%%%%%%%%%%%%%%%%%%%%%%%%%%%%%%%%%%
Next, we perform the calculations for the NNN tunneling terms~$t'$, Eq.~\eqref{eq:Ham-openclosed}, which we can include perturbatively in our description.
I.e. for now we do not assume that the properties of the (sc) and (cc) bound states are affected by NNN tunneling but we only include the terms in small perturbation in~$|t'|<J_\perp$; in Sec.~\ref{sec:truncated-basis} we will consider the case where the open and closed channel are renormalized by~$t'$.
In cuprate materials, the NN and NNN tunneling ratio \mbox{$t/t'<0$~\cite{Ponsioen2019}} is negative, and throughout this study we apply a gauge that fixes~$t>0$.

\begin{figure}[t!]
\includegraphics[width=\linewidth]{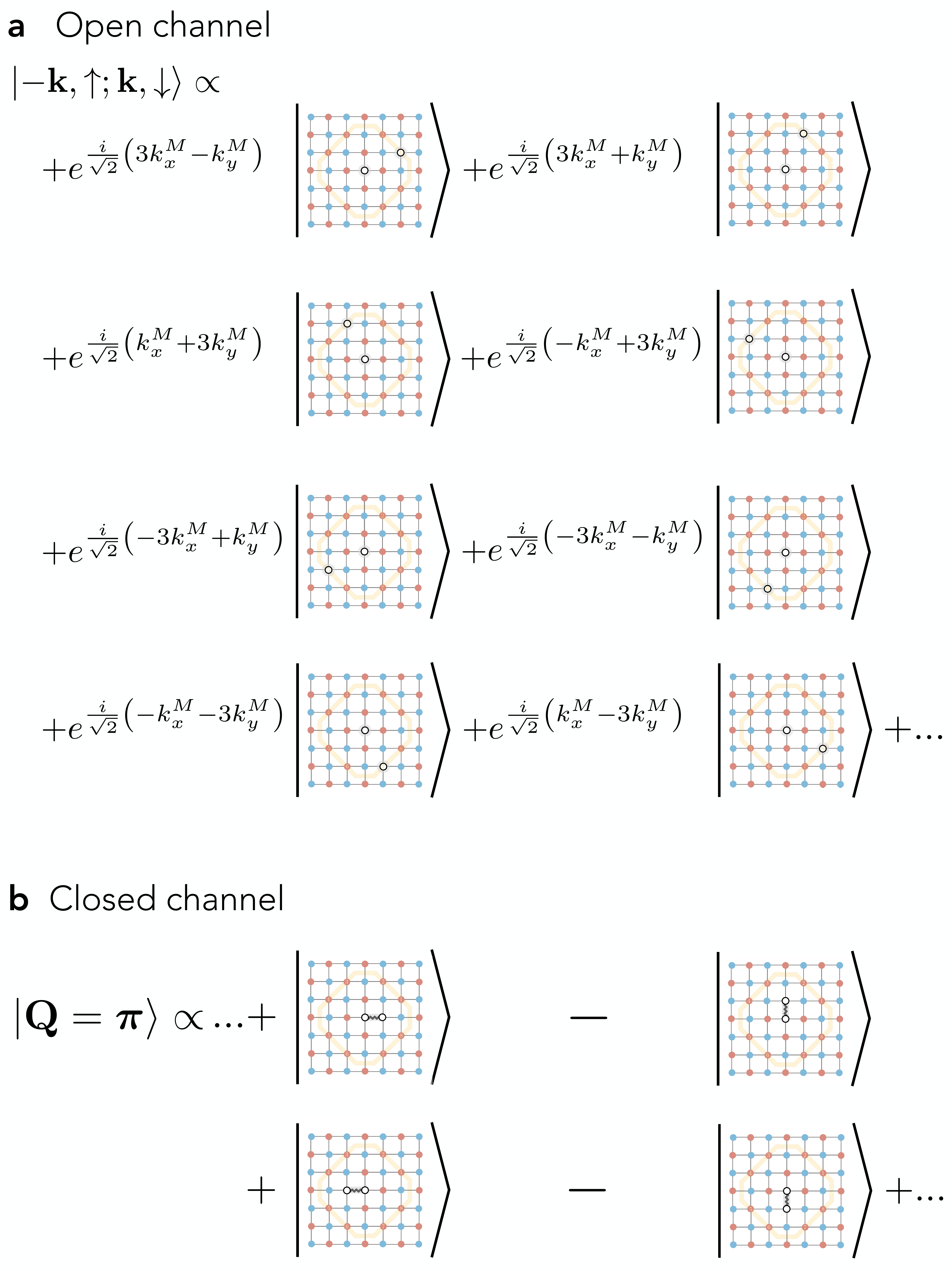}
\caption{\textbf{Shortest string length approximation (SSLA) -- NNN tunneling.} \textbf{a} We show the contributions to the wavefunction of the open channel relevant for NNN tunneling processes, that couple to the closed channel state with~$\ell=1$ in \textbf{b}. The overlaps have to be weighted by momentum-dependent phase factors (here with momenta~$k_x^M,k_y^M$ measured in the $\pi/4$-rotated MBZ basis, see~Eq.\eqref{eq:unitvecs_MBZ}) and symmetry properties of the \mbox{$d$-wave} closed channel have to be taken into account. The shown contributions constitute all eight terms in the SSLA. }
\label{figSSLA-tprime}
\end{figure}

In the following, we evaluate the form factor~$\mathcal{M}_2^{t'}(\mathbf{k})$.
To gain intuition about the processes contributing to NNN tunneling, we illustrate an example in Fig.~\ref{figRecombination}b.
We find that two (sc)'s with strings of length~$\ell_\downarrow=0$ and $\ell_\uparrow$ ($\ell_\uparrow=0$ and $\ell_\downarrow$) can combine to (cc) bound states of length~$\ell=\ell_\uparrow+1$ ($\ell=\ell_\downarrow+1$).
Again, we apply SSLA to calculate the contributions to~$\mathcal{M}_2^{t'}(\mathbf{k})$ for short strings, similar to the procedure described in Sec.~\ref{sec:spinflip}, and we find
\begin{align}
\begin{split}
    \chi^{t'}(\mathbf{k}^M) = &2\Bigg[ \cos{\left(\frac{k_x^M-3k_y^M}{\sqrt{2}}\right)} + \cos{\left(\frac{3k_x^M-k_y^M}{\sqrt{2}}\right)}  \\
    &- \cos{\left(\frac{3k_x^M+k_y^M}{\sqrt{2}}\right)} - \cos{\left(\frac{k_x^M+3k_y^M}{\sqrt{2}}\right)}\Bigg],
\end{split}
\end{align}
where~$\chi^{t'}(\mathbf{k}^M)$ takes the role of~$\chi^{J_\perp}(\mathbf{k}^M)$ as in Eq.~\eqref{eq:formfactor-long}.
The contributing processes are illustrated in Fig.~\ref{figSSLA-tprime}.

To account for long strings, we numerically evaluate the dimensionless form factor~$\mathcal{M}_2^{t'}(\mathbf{k})$, which we show in Fig.~\ref{figSSLA_approx}b together with convergence plots for SSLA.
From the above considerations, we conclude that a (cc)~bound state with string length $\ell_\mathrm{max}=5$, can only couple to magnetic polarons with string length ($\ell_\downarrow=0, \ell_\uparrow \leq 4$) and ($\ell_\downarrow \leq 4, \ell_\uparrow = 0$).
Hence, we again do not encounter Trugman loops or crossings of magnetic polaron strings. 

As for the previous case, we find robust a $d_{x^2-y^2}$ nodal structure, which is caused by the symmetry properties of the closed channel (cc)~bound state.
Further, we note that in our geometric string calculations, the magnitude of the NNN tunneling form factor is large compared to the coupling caused by spin-flip recombinations as can be seen from the scale in Fig.~\ref{figSSLA_approx}.
Hence, we predict notable competition between the two processes even in the perturbative regime~$|t'| \ll J_\perp$, which we discuss in the following.
Importantly, including both spin-flip processes as well as NNN tunneling terms in our model, allows us to analyze the trend of the scattering for doublon (i.e. electron) versus hole doping.

%%%%%%%%%%%%%%%%%%%%%%%%%%%%%%%%%%%%%%%%%%%%%%%%%%%%%
\subsection{Competition between spin-flip and NNN tunneling processes}
\label{sec:competition}
%%%%%%%%%%%%%%%%%%%%%%%%%%%%%%%%%%%%%%%%%%%%%%%%%%%%%
The microscopic Hamiltonian~\eqref{eq:tJ} is formulated in terms of the fermionic operators~$\cd_{\mathbf{j},\sigma}$ describing the underlying electrons in the Fermi-Hubbard model.
However, the magnetic polarons are described in a parton formulation, which requires to introduce the following chargon~$\hat{h}_{\j}$ and spinon~$\hat{s}_{\j,\sigma}$ operators:
\begin{align}
        \c_{\mathbf{j},\sigma} = \begin{cases}  \hd_{\mathbf{j}} \sd_{\mathbf{j},\sigma} ~~~~~~~ \text{for holes} \\  \\ \h_{\mathbf{j}} \s_{\mathbf{j},\sigma} ~~~~~~~ \text{for doublons}\end{cases},
\end{align}
where we distinguish between hole and doublon doping.
Therefore, the underlying microscopic model becomes
\begin{align} \label{eq:Hamiltonian-Schwinger}
\begin{split}
    \H_{\text{$t$-$t'$-$J$}} &=  \pm t \sum_{\ij}  \sum_{\sigma} \hat{\mathcal{P}}\left( \hd_{\mathbf{i}}\h_{\mathbf{j}}\sd_{\mathbf{i},\sigma}\s_{\mathbf{j},\sigma} +\mathrm{h.c.} \right)\hat{\mathcal{P}} \\
    &\pm t' \sum_{\langle\!\ij\!\rangle}\sum_{\sigma}\hat{\mathcal{P}}\left(\hd_{\mathbf{i}}\h_{\mathbf{j}}\sd_{\mathbf{i},\sigma}\s_{\mathbf{j},\sigma}+\mathrm{h.c.} \right)\hat{\mathcal{P}} \\
    &+ J \sum_{\ij} \left( \hat{\mathbf{S}}_\mathbf{i} \cdot \hat{\mathbf{S}}_\mathbf{j}  - \frac{1}{4}\n_\mathbf{i} \n_\mathbf{j}  \right),
\end{split}
\end{align}
for the hole~$(+)$ and doublon~$(-)$ doped case.
In particular, we measure the doping~$\delta$ relative to the half-filled case, $\delta=0$, such that the sign is positive (negative) $\rm{sgn}[\delta]=+1$ ($\rm{sgn}[\delta]=-1$) for hole (doublon) doping.
From the above Hamiltonian, the parton bound state wavefunction can be derived (for~$t'=0$), and we choose a gauge such that all wavefunction amplitudes are real and have a sign structure as follows,
\begin{align}
        \mathrm{sgn}[ \psi_{\rm{sc}}(\ell) ] = \mathrm{sgn}[\phi_{\rm{cc}}(\ell)] = \rm{sgn}[\delta]^\ell,
\end{align}
see Appendix.

\begin{figure*}[t!]
\includegraphics[width=0.9\linewidth]{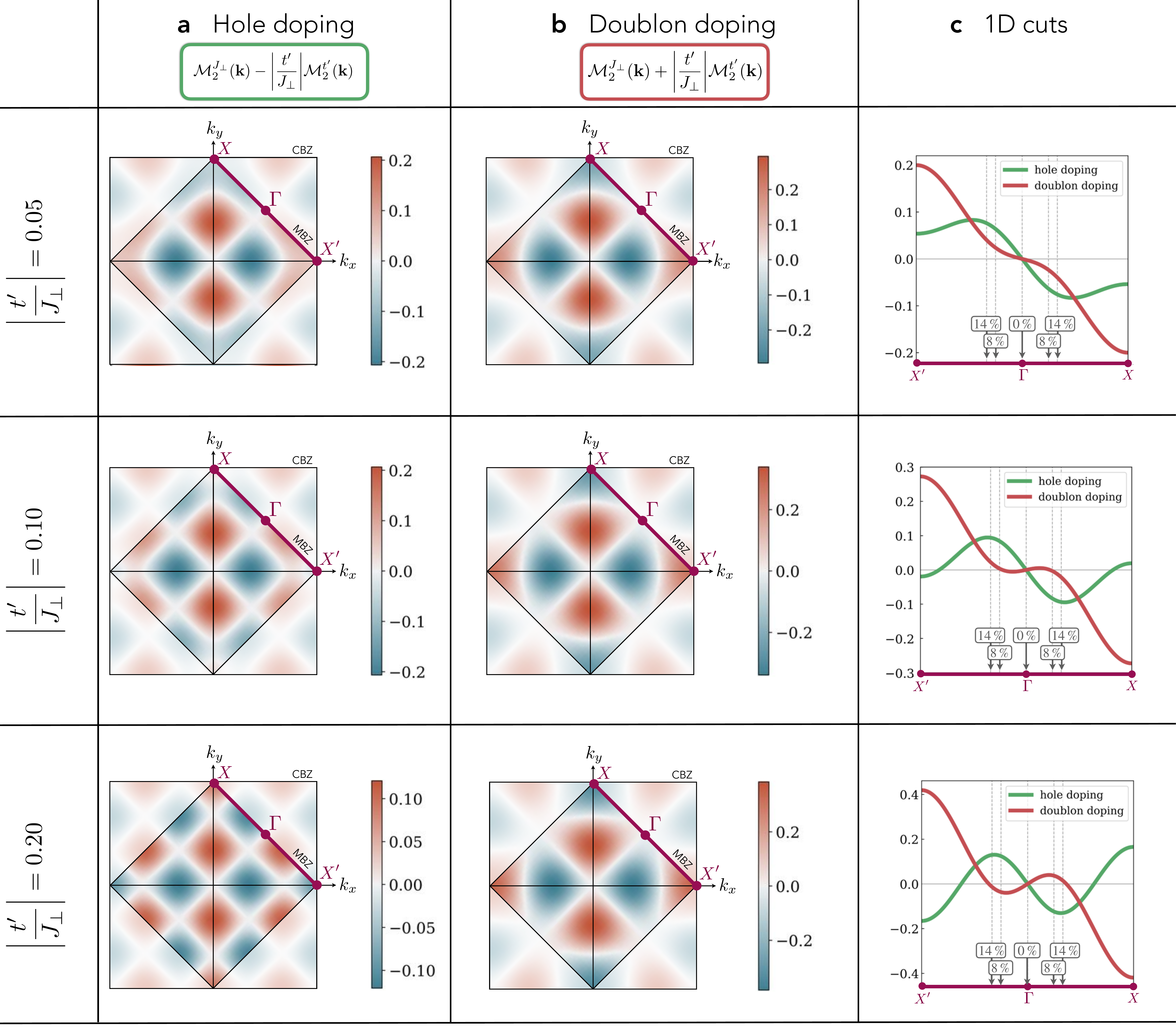}
\caption{\textbf{Combined $J_\perp$ and~$t'$ scattering.} We plot the form facts, Eq.~\eqref{eq:scatteringlength_general}, combining spin-flip and NNN tunneling recombination processes, which enter quadratically in the scattering interaction. \textbf{a} For hole dopants in cuprate compounds, the form factors obtained for spin-flip and NNN tunneling processes are subtracted leading to a constructive interference of the matrix elements around the nodal point~$\mathbf{k}=(\pm \pi/2, \pm \pi/2)$. Here we show the resulting scattering interaction for various parameters $|t'/J_\perp|$. \textbf{b} For doublon doping, the relative sign between the form factors is equal. \textbf{c} We plot 1D cuts along the edge of the Brillouin zone (purple lines in~\textbf{a} and~\textbf{b}). For low doping, the form factors -- and hence scattering amplitude -- gets enhanced (reduced) for hole (doublon) doping. Here, we assume scattering on an elliptical Fermi surface of magnetic polarons with ellipticity~$5.89$~\cite{Martinez1991}. }
\label{figtprime}
\end{figure*}

Next, we consider the doping dependence of the form factors, in which the wavefunction amplitudes enter as products~$\psi_{\rm{sc}}(\ell_\downarrow)\times \psi_{\rm{sc}}(\ell_\uparrow) \times\phi^*_{\rm{cc}}(\ell)$, see Eq.~\eqref{eq:Omega}.
Since~$\ell = \ell_\uparrow + \ell_\downarrow + \Delta \ell$ with~$\Delta \ell = -1,3$ ($\Delta \ell = 1$) for $J_\perp$~($t'$) processes, we find
\begin{subequations} \label{eq:hole-doublon-prefactors}
\begin{align}
    J_\perp \mathcal{M}^{J_\perp}(\mathbf{k})  &\propto \begin{cases} J_\perp\cdot (+1)^{\Delta \ell} = J_\perp ~~~~ \text{for $\delta>0$} \\ \\ J_\perp \cdot (-1)^{\Delta \ell} = -J_\perp ~~ \text{for $\delta<0$} \end{cases}  \\
    t' \mathcal{M}^{t'}(\mathbf{k}) & \propto \begin{cases}  t' \cdot (+1)^{\Delta \ell} = t'  ~~~~~~~\, \text{for $\delta>0$} \\ \\ -t' \cdot (-1)^{\Delta \ell} = t' ~~~~~\, \text{for $\delta<0$} \end{cases} 
\end{align}
\end{subequations}
Therefore, we conclude that in cuprate materials with~$t'/t < 0$ the individual form factors for spin-flip and NNN tunneling recombination processes interfere with a positive (a negative) sign, i.e. the overall form factor is given by ($J_\perp, t>0$)
\begin{align} \label{eq:matrixelement_curlyM_total}
    \mathcal{M}_2^{\rm tot}(\mathbf{k}) = \begin{cases} +J_\perp\mathcal{M}^{J_\perp}_{2}(\mathbf{k}) - |t'| \mathcal{M}^{t'}_{2}(\mathbf{k}) ~~~~~ \text{for $\delta>0$} \\ \\ -J_\perp\mathcal{M}^{J_\perp}_{2}(\mathbf{k}) - |t'| \mathcal{M}^{t'}_{2}(\mathbf{k}) ~~~~~ \text{for $\delta<0$} \end{cases}
\end{align}
The form factors are strongly $\mathbf{k}$-dependent and thus, to meaningful compare the interference effects in the low-doping regime, we need to consider the form factor in the vicinity of the hole pocket's Fermi surface, where~$\mathcal{M}_2^{t'}$ and~$\mathcal{M}_2^{J_\perp}$ have opposite signs, see Fig.~\ref{figtprime}.
As an important result, we find that scattering is enhanced (suppressed) for small hole (doublon) doping.
The form factor determines the strength of the scattering between (sc) charge carriers after integrating out the closed (cc) channel, see Eq.~\eqref{eq:Feshbach}; hence we qualitatively predict that hole doping leads to stronger pairing interactions than doublon doping.

We use the results obtained in geometric string theory, and evaluate the form factor~$\mathcal{M}_2^{\rm tot}(\mathbf{k})$, for various~$|t'/J_\perp|$ in the hole and doublon doped regime, see Fig.~\ref{figtprime}a and b.
The hole pockets are strongly elliptical and thus we expect the scattering of magnetic polarons to occur along the $k_y^M$ axis.
In Fig.~\ref{figtprime}c, we take a 1D cut along the direction in the MBZ and we find that for small doping values, the scattering interaction is strongly enhanced (suppressed) for hole (doublon) doping.
These findings have direct implications for cold atom quantum simulators that can tune NNN tunneling~$t'$ and systematically study hole and doublon doped systems~\cite{Xu2023_Greiner}.

%%%%%%%%%%%%%%%%%%%%%%%%%%%%%%%%%%%%%%%%%%%%%%%%%%%%%
\subsection{Analytical expression of the form factors}
\label{sec:analytical-fit}
%%%%%%%%%%%%%%%%%%%%%%%%%%%%%%%%%%%%%%%%%%%%%%%%%%%%%
The numerically obtained form factors~$\mathcal{M}^{J_\perp}_{2}$ and~$\mathcal{M}^{t'}_{2}$ give insight into the magnitude and nodal structure of the effective interactions.
In the next step, we fit the form factors the analytically obtained functions from SSLA, which allows us to more carefully study the symmetry structure of the pairing interaction, and further enables future (analytical) studies of the effective model, e.g. a BCS mean-field analysis.
In the following, the fitted form factors are denoted by $\tilde{\mathcal{M}}^{J_\perp}_{2}$ and~$\tilde{\mathcal{M}}^{t'}_{2}$, respectively.

We use the results of the numerical calculations including (cc)~states with strings up to length~$\ell_{\rm{max}}=5$ and fit with the momentum dependent functions~$\Gamma_0(\mathbf{k})$ and~$\Gamma_1(\mathbf{k})$ defined below in Eqs.~\eqref{eq:Gamma0} and~\eqref{eq:Gamma1}.
Using this parametrization, we find that the spin-flip form factor is given by 
\begin{align}
    \tilde{\mathcal{M}}_{2}^{J_\perp}(\mathbf{k})=\alpha^{J_\perp} \Gamma_0(\mathbf{k}) + \beta^{J_\perp} \Gamma_1(\mathbf{k})
\end{align}
with~$(\alpha^{J_\perp}, \beta^{J_\perp})=(8.7 \cdot 10^{-2},6.0 \cdot 10^{-2})$.
Analogously, the NNN tunneling form factor becomes 
\begin{align}
\tilde{\mathcal{M}}_{2}^{t'}(\mathbf{k})=\alpha^{t'} \Gamma_0(\mathbf{k})    
\end{align}
with~$\alpha^{t'}=0.55$.
We measure the validity of the fit by evaluating~$\mathrm{err}^{a} = || \tilde{\mathcal{M}}^a-\mathcal{M}^a||_{2}/||\mathcal{M}^a||_{2}$ in the 2-norm.
We find~$\mathrm{err}^{J_\perp}=3.9 \cdot 10^{-3}$ and~$\mathrm{err}^{t'}=4.7 \cdot 10^{-2}$.

Now, we consider the properties of the functions~$\Gamma_0(\mathbf{k})$ and~$\Gamma_1(\mathbf{k})$, which are defined as follows:
\begin{align} \label{eq:Gamma0}
\begin{split}
    \Gamma_0(\mathbf{k})&=\overbrace{\sin{\left(\frac{k_x^M}{\sqrt{2}}\right)}\sin{\left(\frac{k_y^M}{\sqrt{2}}\right)}}^{d_{xy}} \\
    \times&\underbrace{\Bigg\{ -2\sqrt{2} + 4\sqrt{2} \Bigg[ \cos^2{\left(\frac{k_x^M}{\sqrt{2}}\right)} + \cos^2{\left(\frac{k_y^M}{\sqrt{2}}\right)}  \Bigg] \Bigg\} }_{s_0}
\end{split}
\end{align}
\begin{align} \label{eq:Gamma1}
\begin{split}
    \Gamma_1(\mathbf{k})&=\overbrace{\sin{\left(\frac{k_x^M}{\sqrt{2}}\right)}\sin{\left(\frac{k_y^M}{\sqrt{2}}\right)}}^{d_{xy}}\\
    \times &\underbrace{ \bigg[  2 + 4\cos{\left(\sqrt{2}k_x^M\right)} \bigg]\bigg[  2 + 4\cos{\left(\sqrt{2}k_y^M\right)} \bigg]}_{s_1}
\end{split}
\end{align}
The functions have~$d_{xy}$ nodal structure in the $\pi/4$-rotated MBZ basis~$\mathbf{e}_\mu^M$, which translates to a~$d_{x^2-y^2}$ nodal structure in the natural basis~$\mathbf{e}_\mu$ of the CBZ.
Additionally, we conclude that the form factors functions have dependencies on extended $s$-wave channels~$s_0$ and $s_1$.
Further the functions~$\Gamma_p(\mathbf{k})$ with~$p=1,2$ fulfill the following useful orthonormality relations:
\begin{subequations} \label{eq:ONB_integrals}
\begin{align}
    1 &= \frac{1}{2\pi^2}\int_{\mathrm{MBZ}} \Gamma_p(\mathbf{k})\Gamma_p(\mathbf{k})d^2 \mathbf{k}, \\
    0 &= \frac{1}{2\pi^2}\int_{\mathrm{MBZ}} \Gamma_p(\mathbf{k})\Gamma_{\bar{p}}(\mathbf{k}) d^2 \mathbf{k}.
\end{align}
\end{subequations}

\begin{figure}[t!]
\includegraphics[width=\linewidth]{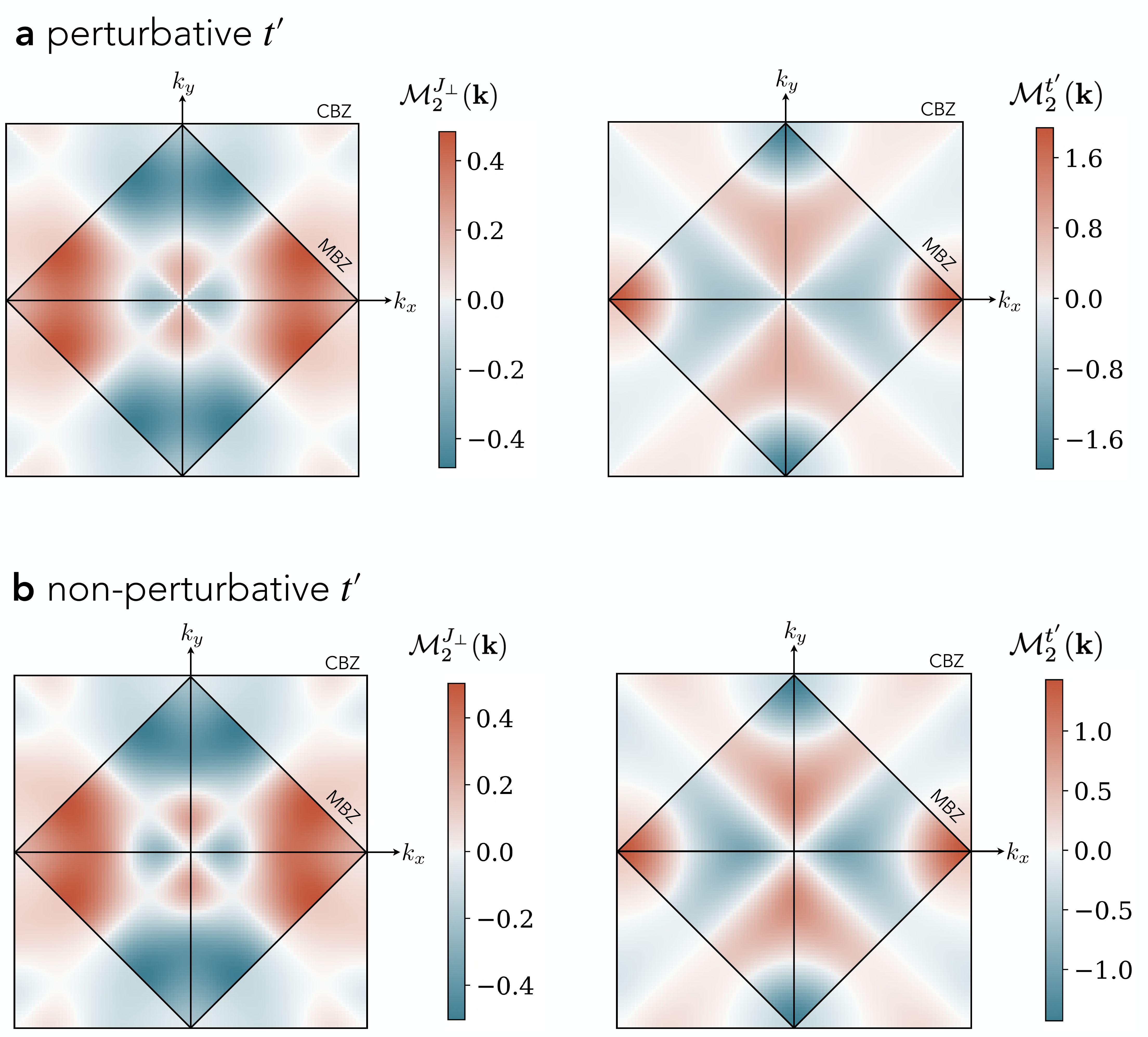}
\caption{\textbf{Scattering form factors} obtained from the refined truncated basis method. The form factors, Eq.~\eqref{eq:scatteringlength_general}, include the full momentum dependence of the (sc) and (cc) wavefunctions. \textbf{a} We plot the form factors~$\mathcal{M}_{2}^{\kappa}$ for~$\kappa=J_\perp$ (left) and~$\kappa=t'$ (right) without including non-perturbative~$t'$ corrections of the meson wavefunctions. \textbf{b} We include the NNN tunneling~$t'$ in the derivation of the (sc) and (cc) wavefunctions using the parameters~$t/J = 3$ and~$t'/t = -0.2$.}
\label{fig_truncated_bare_Jperp_tprime}
\end{figure}
%%%%%%%%%%%%%%%%%%%%%%%%
\section{Refined truncated basis approach}
\label{sec:truncated-basis}
%%%%%%%%%%%%%%%%%%%%%%%%
So far, we have applied a simple Bethe lattice description of the meson bound states, see Fig.~\ref{figBethe}, and we have neglected the momentum dependence of the wavefunction amplitudes. While this had the advantage to obtain analytical expressions for the form factor~$\mathcal{M}_{2}(\mathbf{k})$, Eq.~\eqref{eq:formfactor-long}, predicting quantitative features requires more sophisticate methods.
In the following, we employ a refined truncated basis method, which systematically treats the overcompleteness of the basis states and which allows us to fully take into account the momentum dependencies of the open and closed channel. We find qualitatively excellent agreement with the previous calculations. This demonstrates the robustness of the Feshbach scattering description: the scattering symmetry properties are inherited from the resonant (cc) channel which we capture correctly in the simplified model.

When using the geometric string formalism to describe magnetic polarons (sc), we have so far assumed that the string states $\lbrace\ket{\mathbf{j}_\sigma,\Sigma}\rbrace$ form an orthonormal basis. As mentioned in~\ref{sec:sc-stringtheory}, this is not entirely true since every two trajectories differing by only by a Trugman loop are equivalent and give identical spin configurations (up to global translations) so that the string states form an overcomplete basis set.
The same holds true for the string states of the (cc) $\lbrace \ket{\mathbf{x}_{\mathrm{c}}, \Sigma_{\mathrm{cc}}} \rbrace$, where similar loop effects lead to identical configurations. In addition, the definition of these strings on a Bethe lattice neglects that some trajectories lead to unphysical double occupancies of the chargons and thus overestimates the size of the Hilbert space.

In this section, we will follow~\cite{Bermes2024} and use a more quantitative, refined truncated basis method avoiding the overcompleteness as well as unphysical states, and rigorously including loop effects. To describe the (sc)~meson, we start again from a single hole~$\c_{\mathbf{j},\bar{\sigma}}$ (or electron ~$\c^{\dagger}_{\mathbf{j},\sigma}$) doped into a perfect N\'eel background~$\ket{0}$ and perform a Lee-Low-Pines transformation~\cite{Lee1953} into the co-moving frame of the chargon. This transformation leads to a block-diagonal form in the chargon-momentum basis so that we can compute the wavefunctions for any momentum~$\mathbf{k}$ state. Similar to before, we then consider the chargon motion through a static spin background and construct a truncated basis by applying the NN hopping term~$\hat{H}_{t}$ along sets of bonds consisting of up to~$\ell_{\rm max}$ segments. In contrast to the above, we now do not label the states by the string~$\Sigma$ but by the spin configuration and thus include every physical configuration only once. Since the Ising term $J_z$ gives a confining potential proportional to the number of frustrated bonds, the truncated basis presents a controlled expansion of the Hilbert space relevant for low energy physics.

\begin{figure}[t!]
\includegraphics[width=0.8\linewidth]{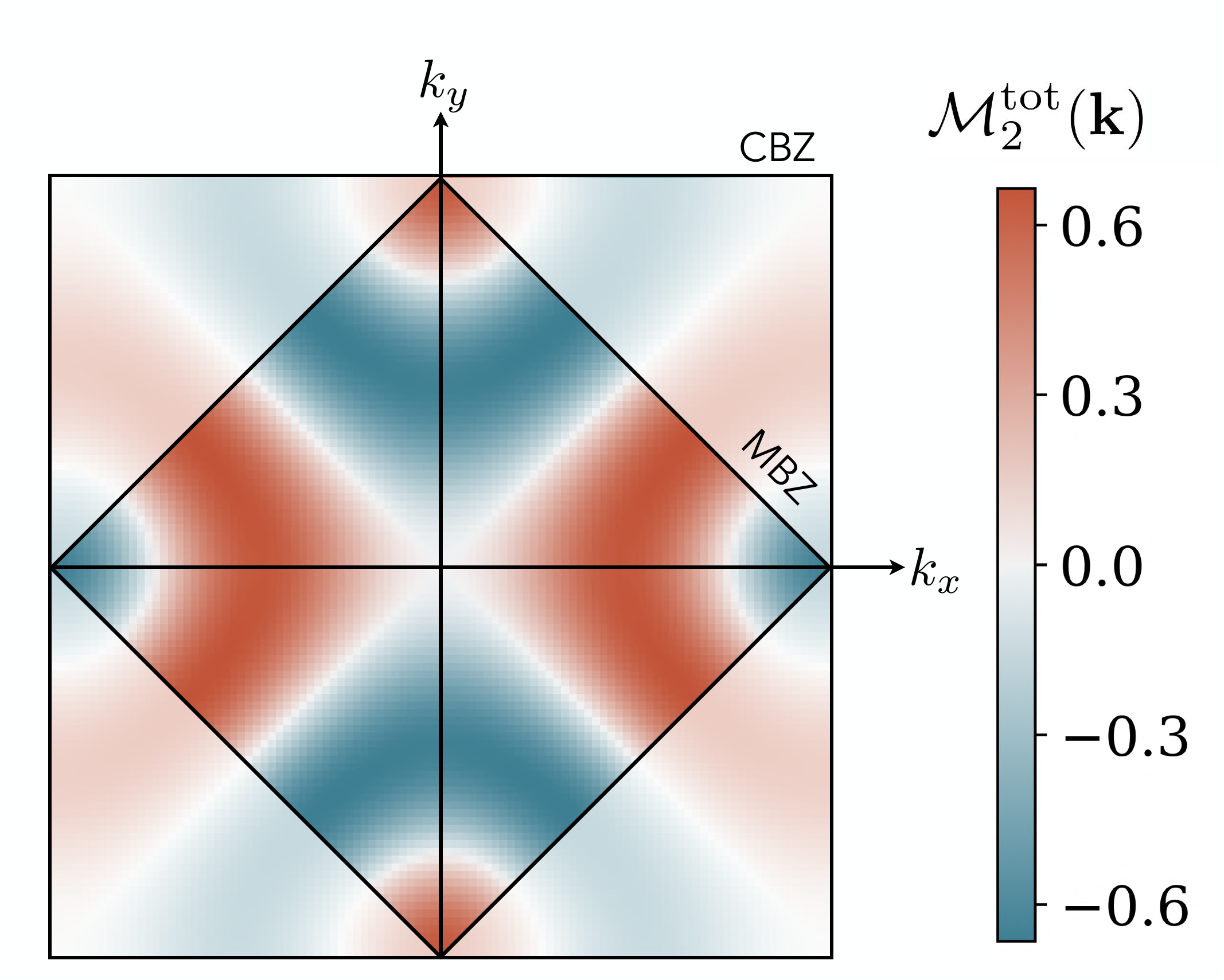}
\caption{\textbf{Combined $J_\perp$ and~$t'$ scattering} obtained from the refined truncated basis method.  We plot the total scattering form factor for hole dopants and realisitc parameters in cuprates, i.e. $t'/t = -0.2$ and $t/J = 3$.}
\label{fig_truncated_combined}
\end{figure}

In order to describe the (cc)~meson, we employ a similar expansion scheme but start from two holes or two doublons doped into a N\'eel background at neighboring sites. Here we assume the dopants to be distinguishable at first and again perform a Lee-Low-Pines transformation into the co-moving frame of the first dopant. As before, we now apply the hopping terms for either of the dopants up to $\ell_{\rm max}$ and build a basis of all distinct and physical configurations. Next, we antisymmetrize the states to account for the fermionic statistics of the indistinguishable chargons.

Having constructed the truncated bases for open channel states~$\mathscr{H}_{\rm open}$ and closed channel states~$\mathscr{H}_{\rm closed}$, we can compute all the matrix elements of the Hamiltonian which do not leave the subspace spanned by the truncated bases. This includes transverse spin fluctuations $J_{\perp}$ and NNN hopping $t'$ which do not break up the geometric strings. Note that we still only include  a subset of all $J_{\perp}$ and $t'$ processes in the system, but we include all the processes which strongly affect the $\rm{(sc)}$ and (cc) wavefunction. The truncated basis description allows us to go beyond the simple Bethe lattice description and to capture non-perturbative $t'$ and $J_\perp$ processes; in particular we include the modifications of the open channel states in the presence of non-zero~$J_\perp, t'$. This leads to a significant momentum dependence of the wavefunction and mixing of the rotational eigensectors away from the C4-invariant momenta. Nevertheless, the dominant contributions to the (sc) wavefunction near the nodal point still have \mbox{$s$-wave} symmetry and at the considered momenta $\mathbf{Q}\,\mathrm{mod}\mathbf{G}^\mathrm{M}=\mathbf{0}$, the (cc) ground state retains its well-defined \mbox{$d$-wave} rotational quantum number.

In order to correctly describe the open ~\mbox{(sc)$^2$} and closed (cc) scattering channels, we have to again take care of the overcompleteness of our parton description. In fact every spin and hole (spin and doublon) configuration contributing to the (cc)~bound state could be written as two individual magnetic polarons~\mbox{(sc)$^2$}. Therefore we adapt the convention, that every spin configuration where the flipped spins form a string connecting the chargons contributes to the (cc) but not to the~\mbox{(sc)$^2$} state.

With this convention we obtain the scattering form factors~$\mathcal{M}^\kappa_2(\mathbf{k})$ [Eq.~\eqref{eq:scatteringlength_general}] shown in Fig.~\ref{fig_truncated_bare_Jperp_tprime}. Here, we truncate the (sc) and (cc) Hilbert spaces at string lengths of $\ell_{\rm max}=8$ and $\ell_{\rm max}^{\rm cc}=10$ to determine the meson wavefunctions. To compute the overlaps~$\mathcal{M}_{2}^{\kappa}$, we used all (sc) states with a string length $\ell < 5$.
The form factors qualitatively agree with the Bethe lattice calculations shown in Fig.~\ref{figSSLA_approx} but have slightly larger values. This is due to the fact, that the truncated basis method includes more possibilities for two individual polarons~\mbox{(sc)$^2$} to recombine into a bound (cc)~pair~\mbox{(cc)}. Furthermore, in the truncated basis method we find that the nodal ring in the form factor of the spin-flip recombination processes~$J_\perp$ moves away from the hole pockets and towards the center of the Brillouin zone.

We compare our calculations by (i) not including the~$J_\perp$ and $t'$~terms in the meson's wavefunctions, see Fig.~\ref{fig_truncated_bare_Jperp_tprime}a, and (ii) fully including the $t/J_\perp=3$ and $t'/t=-0.2$ dependencies of the open and closed channels, see Fig.~\ref{fig_truncated_bare_Jperp_tprime}b, using typical parameters of hole doped cuprate superconductors~\cite{Ponsioen2019}.
We find that the non-perturbative corrections of the meson wavefunction leads to very minor differences justifying the perturbative treatment of $t'$-processes. Note that including the~$J_\perp$ processes is crucial to obtain the hole pockets of the magnetic polarons; a comparison between the geometric string model to numerical DMRG studies of the \mbox{$t$-$J$} model is discussed in Ref.~\cite{Grusdt2019}.

In Fig.~\ref{fig_truncated_combined}, we show the combined scattering form factor $\mathcal{M}_{2}^{\textrm{tot}}=\mathcal{M}_{2}^{J_\perp} - |\frac{t'}{J_\perp}|\mathcal{M}_{2}^{t'}$ for hole dopants, see Eq.~\eqref{eq:matrixelement_curlyM_total}, including NNN tunneling~$t'$ non-perturbatively. In the vicinity of the hole pocket's Fermi surface, the NNN tunneling and spin-flip processes interfere constructively leading to sizable scattering form factors.

%%%%%%%%%%%%%%%%%%%%%%%%%%%%%%%%%%%%%%%%%%%%%%%%%%%%%
\section{Effective Hamiltonian}
\label{sec:effModel}
%%%%%%%%%%%%%%%%%%%%%%%%%%%%%%%%%%%%%%%%%%%%%%%%%%%%%
So far, we have derived scattering interactions~$V_{\mathbf{k},\mathbf{k'}} \propto \mathcal{M}^*(\mathbf{k})\mathcal{M}(\mathbf{k}')$ of two magnetic polarons in terms of the form factors and for zero-momentum pairs. 
Next, we want to take another significant step by promoting our two-body problem to a many-body theory.
Our effective theory leads us to an effective model to describe \mbox{$d$-wave} superconductivity in the low-doping and strong-coupling regime arising from a Fermi sea of magnetic polarons.

The effective model derived below is valid in a regime, when (i)~the antiferromagnetic background has sufficiently long-ranged correlations, $\xi_{\mathrm{AFM}} \gg a$, and (ii)~the density of magnetic polarons, i.e. charge carriers, is low, and (iii)~the system can be described by low-energy scattering, i.e. at low temperatures~$k_B T \ll J_\perp, t'$.
The requirement~(ii) ensures that the internal structure of the meson-like bound state does not have to be adjusted due to substantial overlaps of the (sc) or (cc) wavefunctions; for the former this is expected to play a role at doping~$\delta \gtrsim 20\%$~\cite{Chiu2019,Koepsell2021}.

The coupling matrix elements (or form factors) between the open and closed channel, see Sec.~\ref{sec:ScattLength}, allow us to integrate out the closed channel and derive an effective Hamiltonian for the magnetic polarons with pairwise scattering in momentum space given by~\cite{Homeier2023Feshbach}
\begin{align} \label{eq:effHam}
    \H_\mathrm{eff} &= \H_\mathrm{open} + \H_{\mathrm{int}}, \\
    \H_{\mathrm{int}} &= \sum_{\mathbf{k},\mathbf{k}'} V_{\mathbf{k},\mathbf{k}'} \pd_{-\mathbf{k},\uparrow} \pd_{\mathbf{k},\downarrow} \p_{-\mathbf{k}',\uparrow}\p_{\mathbf{k}',\downarrow}
\end{align}
with the open channel Hamiltonian describing weakly interacting magnetic polarons, see Eq.~\eqref{eq:Hopen}.
For low doping, the Fermi surface forms two hole pockets around the dispersion minima~$(\pm \pi/2, \pm\pi/2)$.
Further, the effective interaction matrix elements~$V_{\mathbf{k},\mathbf{k}'}$, Eq.~\eqref{eq:scatteringlength_general}, arise from the emergent \mbox{$d$-wave} Feshbach resonance between~$\rm{sc}^2$ and (cc) mesons with total momentum~$\mathbf{Q}=\mathbf{0}$.

The above derived analytical expressions of the form factors allow us to give a closed form of the attractive two particle scattering interaction,
\begin{align}
    V_{\mathbf{k},\mathbf{k}'} &= \frac{1}{\mathcal{V}} \left[ g_0(t') \Gamma_0(\mathbf{k}') + g_1 \Gamma_1(\mathbf{k}') \right]\left[ g_0(t') \Gamma_0(\mathbf{k}) + g_1 \Gamma_1(\mathbf{k}) \right],
\end{align}
where~$\mathcal{V} = L^2/2$ is the volume of the magnetic lattice.
Further, we have defined the coupling constants
\begin{align}
    g_0(t') &:= \sqrt{\frac{J^2_\perp}{2\Delta E}} \,\left( \alpha^{J_\perp} + \frac{t'}{J_\perp}\alpha^{t'} \right) \\
    g_1 &:= \sqrt{\frac{J^2_\perp}{2\Delta E}}\, \beta^{J_\perp},
\end{align}
where~$\Delta E$ denotes the bare energy splitting between the open and closed channels, see Section~\ref{sec:cc-stringtheory}.
This energy splitting strongly determines the couplings strength between the two channels. While we introduce $\Delta E$ as a free tuning parameters in the meson scattering model, microscopic couplings such as extended Hubbard interactions eventually determine the proximity to the Feshbach resonance of a given model~\cite{Homeier2023Feshbach,Lange2023,Lange2024Lang} and may allow to tune the effective interaction strength between charge carriers in solids or cold atom experiments.

In the following, we normalize the factors of the scattering interactions,
\begin{align}
    \Gamma(\mathbf{k}) = \mathcal{N}^{-1/2} \left[ g_0(t') \Gamma_0(\mathbf{k}') + g_1 \Gamma_1(\mathbf{k}') \right]
\end{align}
such that $(2\pi^2)^{-1/2}\int_{\mathrm{MBZ}}\Gamma^2(\mathbf{k}) d^2\mathbf{k} = 1$. Using Eqs.~\eqref{eq:ONB_integrals}, this gives~$\mathcal{N} = g^2_0(t')+g_1^2$.
This yields the final expression for the scattering interaction
\begin{subequations}
\begin{align}
     V_{\mathbf{k},\mathbf{k}'} &= \frac{g}{\mathcal{V}} \Gamma(\mathbf{k})\Gamma(\mathbf{k}'), \\
     g &:= g^2_0(t')+g_1^2,
\end{align}
\end{subequations}
such that we arrive at a BCS Hamiltonian with highly anisotropic pairing interactions, which can lead to an instability of the Fermi surface of magnetic polarons, i.e. around the small, elliptical hole pockets at the nodal point.

%%%%%%%%%%%%%%%%%%%%%%%%%%%%%%%%%%%%%%%%%%%%%%%%%%%%%
\subsection{BCS mean-field analysis}
\label{sec:BCS-MF}
%%%%%%%%%%%%%%%%%%%%%%%%%%%%%%%%%%%%%%%%%%%%%%%%%%%%%
We treat the Hamiltonian~\eqref{eq:effHam} using a standard BCS mean-field ansatz in order to analyze the symmetries of the BCS pairing gap~$\Delta(\mathbf{k})$.
We define the \mbox{$d$-wave} BCS mean-field order parameter
\begin{align}
    \Delta(\mathbf{k}) &= \frac{1}{\mathcal{V}}\sum_{\mathbf{k}'}\Gamma(\mathbf{k}) \Gamma(\mathbf{k}') \langle \p_{-\mathbf{k}',\uparrow}\p_{\mathbf{k}',\downarrow} \rangle,
\end{align}
which is the gap equation that has to be fulfilled self-consistently in accordance with the BCS mean-field Hamiltonian,
\begin{align} \label{eq:H_MF_appendix}
\begin{split}
    \H_\mathrm{MF} = &\sum_{\mathbf{k}} \left[ \epsilon_\mathrm{sc}(\mathbf{k}) - \mu \right] \pd_{\mathbf{k},\sigma}\p_{\mathbf{k},\sigma} \\
    + g&\sum_{\mathbf{k}}\Delta(\mathbf{k})  \big[\pd_{-\mathbf{k},\uparrow} \pd_{\mathbf{k},\downarrow} +\mathrm{H.c.} \big].
\end{split}
\end{align}
From an ansatz for the pairing gap,~$\Delta(\mathbf{k})=\Delta \Gamma(\mathbf{k})$, it immediately follows that the pairing gap has the same symmetry and nodal structure as the form factors shown in Fig.~\ref{figtprime}, i.e.~$\Delta(\mathbf{k}) \propto \mathcal{M}_2^{\rm tot}$.

The magnitude of the gap and hence the mean-field transition temperature has to be determined self-consistently and strongly depends on (i) the interaction strength $g \propto \Delta E_2^{-1}$ as well as (ii) the Fermi energy~$E_F$.
In particular, in the low-doping regime these two energy scales are competing, which may lead to a non-mean field character of the phase transition as the temperature is lowered going beyond the scope of this study.
We conclude that in the proposed Feshbach scenario~\cite{Homeier2023Feshbach} a \mbox{$d$-wave} superconductor can be established and is the leading order instability of a magnetic polaron metallic state recently observed in Ref.~\cite{Kurokawa2023}.
However, the details of the BCS state, e.g. the magnitude of the pairing gap, strongly depends on the bare energy splitting between the open channel and closed \mbox{$d$-wave} channel~$\Delta E_2^{-1}$, which requires future numerical and experimental studies, such as spectroscopy of the (cc)~meson.

Before we discuss direct spectroscopic signatures of the tightly-bound (cc) state in Sec.~\ref{sec:ARPES}, we emphasize that the Bogoliubov quasiparticle dispersion, $E(\mathbf{k}) = \sqrt{ \left[\epsilon_{\mathrm{sc}}(\mathbf{k})-\mu\right]^2 + \Delta^2(\mathbf{k})}$, is an indirect probe of our scenario accessible in single particle spectroscopy, e.g. ARPES.
The Bogoliubov dispersion is linked to the form factors derived in Sec.~\ref{sec:ScattLength}, for which we find characteristic features such as an nodal ring-like structure, see Fig.~\ref{figtprime}. 
More directly, the momentum dependence of the superconducting pairing gap~$\Delta(\mathbf{k}) \propto \Gamma(\mathbf{k}) \propto \mathcal{M}(\mathbf{k})$ can be analyzed as we discuss next.

\begin{figure*}[t!]
\includegraphics[width=\linewidth]{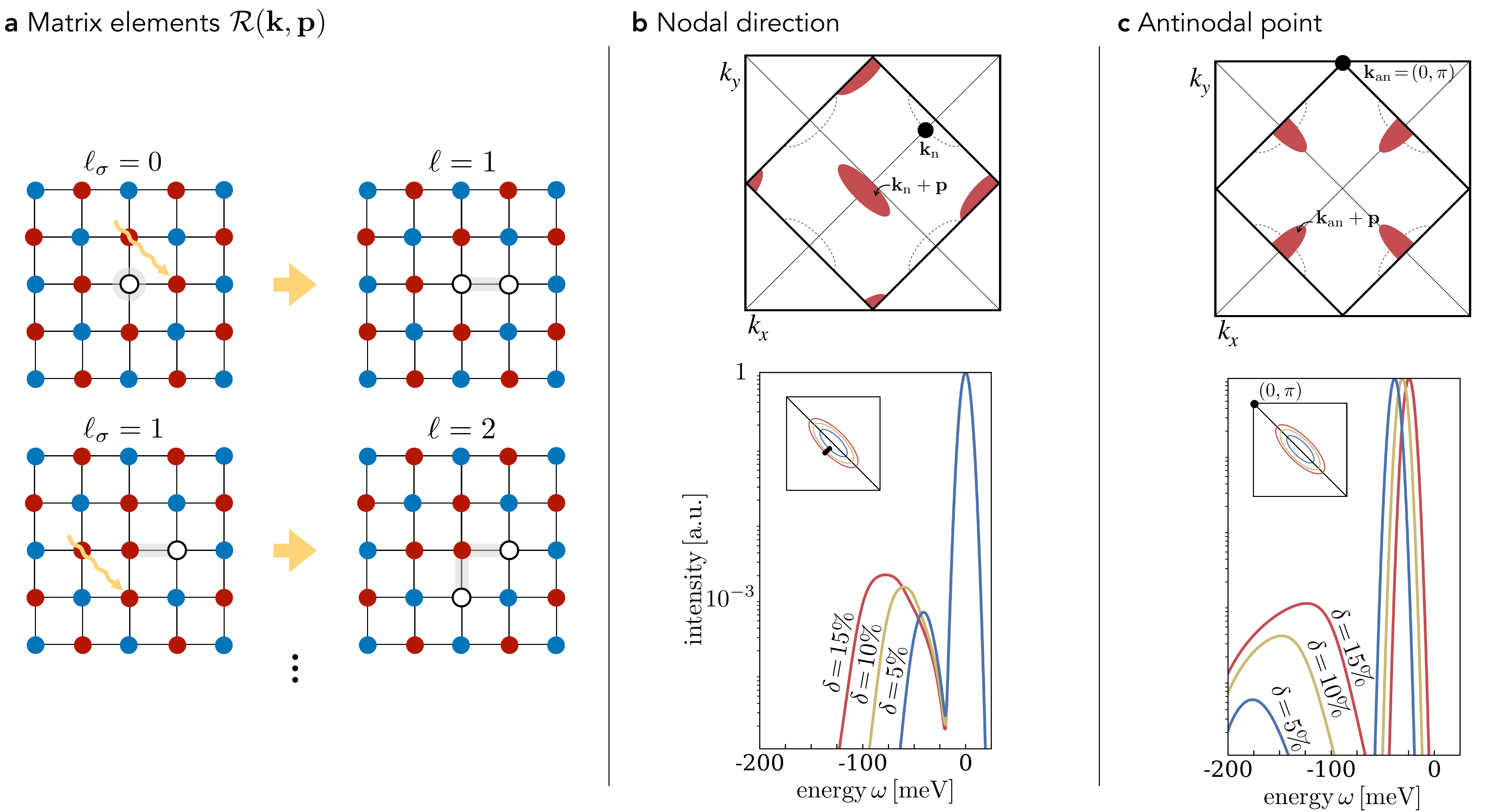}
\caption{\textbf{Single-hole ARPES.} \textbf{a} In single-hole ARPES, a rare but possible process describes the removal of an electron close an existing (sc)~meson. The matrix elements are calculated by expanding in string length states; we include strings up to length~$\ell_\sigma \leq 1$, as shown for two examples. We show the spectral signal \textbf{b} in the nodal direction, i.e. $\mathbf{k}_{\rm n}$ at the Fermi surface of the hole pocket and on the diagonal of the CBZ, and \textbf{c} at the antinodal point~$\mathbf{k}_{\rm an}=(0,\pi)$.  Top panels: The two-body processes, involving the removal of a (sc) and the addition of a (cc) meson, requires to convolve the (sc) dispersion~$\varepsilon_{\rm sc}$ at the hole pocket around~$\mathbf{p}=(\pm\pi/2,\pm\pi/2)$ (dotted region) with the shifted (cc) dispersion~$\varepsilon_{\rm cc}$ at~$\mathbf{k}_{\rm (a)n} + \mathbf{p}$ (red region). The gradient of the (cc) dispersion in the red regions, together with the matrix elements~$\mathcal{R}(\mathbf{k},\mathbf{p})$ determines the width of the spectral signal. Bottom panels: We plot the spectral signal for~$J=130\,\mathrm{meV}$,~$T=1.3\,\mathrm{K}$,~$\Delta E_{2}=20\,\mathrm{meV}$ and various hole dopings~$\delta=5\,\%,10\,\%,15\,\%$. The large peaks corresponds to the quasiparticle peak of the (sc)~mesons. We find a weak and broad feature below the Fermi surface associated with the (cc) bound state and with an energy onset at~$\Delta E_{2}$ (note the logarithmic scale, however).
 }
\label{figARPES-1hole}
\end{figure*}
%%%%%%%%%%%%%%%%%%%%%%%%
\section{ARPES signatures}
\label{sec:ARPES}
%%%%%%%%%%%%%%%%%%%%%%%%

An established experimental technique to probe the electronic structure of materials is ARPES~\cite{Sobota2021}.
The $\rm{(sc)}^2$ Feshbach scattering channels described in this article rely on the existence of the tightly-bound bosonic (cc)~pair~\cite{Homeier2023Feshbach}.
The properties of the (sc) mesons, or magnetic polaron, are directly accessible in conventional ARPES experiments, where a one-photon-in-one-electron-out process is considered.
This process corresponds to the creation of a single hole excitation in the material and thus strongly couples to the individual (sc)~channel.

Probing the (cc)~bound state, however, is much more challenging because it involves two correlated holes. In ARPES, the (cc)~bound state can be probed by (i) a process, in which an additional single hole couples to an already existing (sc) meson and forms a (cc) meson, and by (ii) correlated two-photon-in-two-electron-out processes (cARPES). 
In the following, we calculate the matrix elements for both processes and we find that the process described in (i) only couples very weakly to the (cc)~channel.

%%%%%%%%%%%%%%%%%%%%%%%%
\subsection{Single hole ARPES}
\label{sec:ARPES-1hole}
%%%%%%%%%%%%%%%%%%%%%%%%
In the low but finite doping regime, we describe the fermionic charge carriers as magnetic polarons with dispersion relation~$\varepsilon_{\rm sc}(\mathbf{k})$ in a Fermi sea, see Eq.~\eqref{eq:sc-dispersion}. In a photoemission process, a hole with momentum~$\mathbf{k}$ and energy~$\omega$ can be created. If this additional hole is in the vicinity of an already existing spinon~(s) with momentum~$\mathbf{p}$, which is bound to a chargon~(c), they can recombine into a (cc)~bound state with momentum~$\mathbf{k}+\mathbf{p}$ and energy~$\varepsilon_{\rm cc}(\mathbf{k}+\mathbf{p})$ while leaving behind a hole-like ${\rm (\overline{sc})}$ excitation.
The (cc) dispersion for the $d$-wave channel in the \mbox{$t$-$J$}~model cannot be calculated in the simple geometric string picture but has been extracted from DMRG calculations~\cite{Bohrdt2023_dichotomy},
\begin{align}
\varepsilon_{\rm cc}(\mathbf{p})=-J (\cos(p_x)+\cos(p_y)-2]+\Delta E_{2},
\end{align}
showing the relatively light mass~$\propto 1/J$ of the bi-polaronic state.

The corresponding ARPES signal is determined by a convolution of the matrix elements with the Fermi sea of magnetic polarons given by
\begin{align}
\begin{split} \label{eq:ARPES}
    A_{\overline{\rm sc}+{\rm cc}}(\mathbf{k},\omega) = &\int \frac{d^2\mathbf{p}}{(2\pi)^2}~ n^{\rm F}_T ( \varepsilon_{\rm{sc}}(\mathbf{p}) - \mu ) ~ \mathcal{R}(\mathbf{k},\mathbf{p}) \\
    &\times ~ \delta(\omega + \varepsilon_{\rm{sc}}(\mathbf{p}) - \mu - \varepsilon_{\rm{cc}}(\mathbf{k+\mathbf{p}})  ).
\end{split}
\end{align}
Here,~$n^{\rm F}_T(\varepsilon)$ is the Fermi-Dirac distribution at temperature~$T$ and we assume sufficiently low temperature such that the thermal occupation of the (cc)~channel can be neglected; further $\mathcal{R}(\mathbf{k},\mathbf{p})$ are coupling matrix elements, see Fig.~\ref{figARPES-1hole}a.
Note that the single-hole APRES process gives rise to a two-particle continuum and therefore a broad spectral feature.

The matrix elements~$\mathcal{R}(\mathbf{k},\mathbf{p})$ are evaluated by expanding the (cc) and (sc) wavefunctions in the string length basis analogously to the calculation of the scattering form factor in Sec.~\ref{sec:ScattLength}. Moreover, we approximate the meson wavefunction to be momentum independent and assume the ground state wavefunctions in the respective channel. 
Since the meson wavefunctions only extend across a few lattice sites, we only consider (sc) strings up to length~$\ell_\sigma = 1$, which couple to (cc) contributions of length~$\ell=1,2$.
Using our approximation, we find
\begin{widetext}
\begin{align} \label{eq:matrixelement_R}
\begin{split}
    \mathcal{R}(\mathbf{k},\mathbf{p}) &=\sqrt{2}\psi_{\rm sc}(\mathbf{k},\ell_\sigma = 0) \phi_{\rm cc}(\mathbf{p},\ell=1)\big[ \cos(k_x) - \cos(k_y) - \cos(p_x) + \cos(p_y)  \big] \\
    &+\sqrt{2}\psi_{\rm sc}(\mathbf{k},\ell_\sigma = 1) \phi_{\rm cc}(\mathbf{p},\ell=1)\big[ \cos(k_x + p_x) - \cos(k_y+p_y)\big] \\
    &+\sqrt{2}\psi_{\rm sc}(\mathbf{k},\ell_\sigma = 1) \phi_{\rm cc}(\mathbf{p},\ell=2)\big[  \cos(k_x + k_y + p_x) - \cos(k_x + k_y + p_y) \\
    &+\cos(k_x - k_y + p_x)-  \cos(-k_x + k_y + p_x) - 2 \cos(p_x) + 2\cos(p_y)\big] 
\end{split}
\end{align}
\end{widetext}
with momenta defined in the CBZ. Moreover, the matrix element~\eqref{eq:matrixelement_R} only contains contributions from $s$-wave (sc)~states and $d$-wave (cc)~states, despite small mixing with~$p$-wave (sc)~state at the nodal point.
Nevertheless, we expect the approximation to be valid in the very low doping regime, as we confirmed using the systematic truncated basis approach from Sec.~\ref{sec:truncated-basis}.
For finite doping, the (sc) and (cc)~wavfunctions will be adapted even further due to interactions and reduced AFM correlations; we neglect such effects to~$\mathcal{R}(\mathbf{k},\mathbf{p})$ in the following.

We use the matrix element~\eqref{eq:matrixelement_R} to calculate the spectral weight in different regions~$\mathbf{k}$ of the Brillouin zone.
In particular, the two-body origin of the spectral feature leads to a contribution of the (cc)~state at~$\mathbf{k}+\mathbf{p}$, where~$\mathbf{p}$ are the occupied momenta of (sc)'s in the hole pocket, see Fig.~\ref{figARPES-1hole}b and c (top).
The confined meson bound states are assumed to exist in the low-doping regime, where sufficient AFM correlation are present. Since the density of (sc) mesons is low in this regime and the matrix elements are small, we predict only very weak spectroscopic signatures with a maximum peak height of about~$10^{-3}$ relative to the quasiparticle peak of the (sc)'s.

In Fig.~\ref{figARPES-1hole}b and c (bottom), we analyze the ARPES signal for typical cuprate parameters, various hole dopings~$\delta$ and for two common momenta in the Brillouin zone, i.e. at the Fermi surface in the nodal direction~$\mathbf{k}_{\rm n}$ and at the antinodal point~$\mathbf{k}_{\rm an}$. We assume a bare open-closed channel energy difference of~$\Delta E_2= 20\,{\rm meV}$.
Along the nodal direction, see Fig.~\ref{figARPES-1hole}b, we find a pronounced peak with a full-width-half-max (FWHM) around $15\,{\rm meV}$. While the onset of the signal, at $\Delta E_2$ below $E_F$, is a fit parameter the FWHM only depends on the value of $J$ without further fit parameters. Along the antinodal direction, see Fig.~\ref{figARPES-1hole}c, we find a much broader hump followed by a dip between $15$ to $25\,{\rm meV}$. The onset of this feature, defined by the dip, remains at an energy scale $\Delta E_2 \simeq 20\,{\rm meV}$ as observed in nodal direction. 

In underdoped cuprates other, more pronounced peak-dip-hump features have previously been observed~\cite{Zhou2002}, dimming the prospect of detecting the weak signal we find in Fig.~\ref{figARPES-1hole}.

%%%%%%%%%%%%%%%%%%%%%%%%
\subsection{Coincidence ARPES}
\label{sec:ARPES-2hole}
%%%%%%%%%%%%%%%%%%%%%%%%
Alternatively, it was proposed in Refs.~\cite{Bohrdt2023_dichotomy,Homeier2023Feshbach} to use correlated two-hole spectroscopy, which is highly sensitive to the existence of the (cc)~channel, and can be realized by cARPES~\cite{Berakdar1998,Mahmood2022,Su2020}.
We consider processes with two in-coming photons with momentum~$\mathbf{K}$ and two out-going photoelectrons with momentum~$\mathbf{K} \pm \mathbf{k}$, and we discuss the matrix elements~$\mathcal{C}(\mathbf{k})$ for the shortest string length~$\ell_{\rm }=1$.
For the specific momenta we consider, the process has no momentum transfer to the sample and allows us to probe the (cc)~channel at~$\mathbf{Q}=\mathbf{0},\bm{\pi}$.
The cARPES matrix elements for short strings is calculated to be
\begin{align}
    \mathcal{C}(\mathbf{k}) \propto |\psi_{\rm sc}(\mathbf{k},\ell_{\sigma}=0)|^2 \phi_{\rm cc}(\mathbf{Q},\ell_{\rm cc}=1) \big[ \cos(k_x) \pm \cos(k_y) \big]
\end{align}
for momenta in the CBZ and the $s$-wave ($+$) and $d$-wave ($-$) channel.
Therefore, we find a sizable lower bound for matrix elements of the (cc)~channel in cARPES with distinct symmetry features such as a nodal structure of the $d$-wave pair inherited from its $m_4$~eigenvalue.

%%%%%%%%%%%%%%%%%%%%%%%%
\section{Summary and Outlook}
\label{sec:summary-outlook}
%%%%%%%%%%%%%%%%%%%%%%%%
We have theoretically developed a scattering theory for low-energy spinon-chargon~(sc) and chargon-chargon (cc) meson excitations of doped AFMs.
The various internal excitations of the individual charge carriers give rise to a multichannel description -- one channel for each set of quantum numbers -- and Feshbach resonances if a pair of ${\rm (sc)}^2$ recombines into an excited (cc)~state.
In Ref.~\cite{Homeier2023Feshbach}, it has been suggested that in cuprate superconductors a resonant $d$-wave bi-polaronic state could lead to strong attractive pairing; in this work, we have analyzed the $d$-wave scattering channel in greater detail and provided ab-inito calculations based on a truncated basis method of confined strings in a N\'eel background.
Our method allows us to discuss further implications of the Feshbach scenario, such as comparing hole and doublon doping, or calculating ARPES matrix element relevant for experimental tests of the multichannel perspective.
Our findings enable future studies of a potential BEC-BCS crossover in this model, non-Fermi liquid behavior or a quantitative BCS mean-field and Eliashberg analysis of the proposed model, in order to quantitatively study the phase diagram of lightly doped quantum magnets. These studies may be extended by including triplet scattering channels and finite momentum Cooper pairs using our developed formalism.

The Feshbach perspective, leading to mediated strong interactions in many-body systems, has recently gained attention in 2D heterostructures~\cite{Sidler2016,Schwartz2021,Kuhlenkamp2022} and strongly-correlated electrons~\cite{Slagle2020,Crepel2021,Lange2023,Yang2023,Milczewski2023,Zerba2023}.
Whether the Feshbach hypothesis is realized in cuprate superconductors relies on the existence of the (cc)~channel. As we calculate explicitly, coincidence ARPES spectroscopy offers the possibility to detect the closed channel (cc)~states.

Moreover, indirect probes or probes in related systems could be used to search for related scenarios in doped quantum magnets. 
These include pump-probe experiments~\cite{Homeier2023Feshbach}, spectroscopy in antiferromagnetic bosonic $t$-$J$ models~\cite{Homeier2023}, or real-space correlation measurements in ultracold atoms~\cite{Bohrdt2021Review}. The latter platforms are highly tunable, clean systems, which e.g. enables to systematically include $t'$ terms~\cite{Xu2023_Greiner}.

%%%%%%%%%%%%%%%%%%%%%%%%
\section*{Acknowledgments}
%%%%%%%%%%%%%%%%%%%%%%%%
We thank Annabelle Bohrdt, Eugene Demler, Hannah Lange and Krzysztof Wohlfeld for inspiring discussions.  -- This research was funded by the Deutsche Forschungsgemeinschaft (DFG, German Research Foundation) under Germany's Excellence Strategy -- EXC-2111 -- 390814868 and has received funding from the European Research Council (ERC) under the European Union’s Horizon 2020 research and innovation programm (Grant Agreement no 948141) — ERC Starting Grant SimUcQuam. L.H. was supported by the Studienstiftung des deutschen Volkes.

%%%%%%%%%%%%%%%%%%%%%%%%%%%%%%%%%%%%%%%%
% INSERT BIBLIOGRAPHY LOCATION HERE:

%merlin.mbs apsrev4-1.bst 2010-07-25 4.21a (PWD, AO, DPC) hacked
%Control: key (0)
%Control: author (8) initials jnrlst
%Control: editor formatted (1) identically to author
%Control: production of article title (0) allowed
%Control: page (1) range
%Control: year (1) truncated
%Control: production of eprint (0) enabled
%

\newpage~
\newpage~
\onecolumngrid
\appendix

%%%%%%%%%%%%%%%%%%%%%%%%%%%%%%%%%%%%%%%%%%%%%%%%%%%%%
\section*{Appendix}
%%%%%%%%%%%%%%%%%%%%%%%%%%%%%%%%%%%%%%%%%%%%%%%%%%%%%
The string construction is based on a mapping from the \mbox{$t$-$J$} Hilbert space onto the string Hilbert space~$\mathscr{H} = \mathscr{H}_{\rm open}\oplus \mathscr{H}_{\rm closed}$. In this procedure, the matrix elements of the Hamiltonian have to be determined in a consistent way in order to correctly take care of the fermionic statistics of the underlying constituents.
In the following, we consider hole doping and use a Schwinger bosons representation of the underlying electrons, i.e.~$\c_{\j,\sigma} = \hat{s}_{\j,\sigma}\hd_{\j}$, with the local number constraint~$\sum_{\sigma}\hat{s}^\dagger_{\j,\sigma}\hat{s}_{\j,\sigma} + \hd_{\j}\h_{\j}$. Further, we will use the Hamiltonian as defined in the main text in Eq.~\eqref{eq:Hamiltonian-Schwinger}. \\

\emph{Spinon-chargon states.}
In order to choose the string wavefunction amplitudes~$\psi_{\rm sc}$ positive and real, we define the string states as
\begin{subequations}
\begin{align}
    \ket{\j_\sigma, \Sigma=0} &= \c_{\j,\bar{\sigma}}\ket{0} = \hat{s}_{\j,\bar{\sigma}}\hd_{\j}\ket{0} \\
    \ket{\j_\sigma, \Sigma' = \Sigma+\mathbf{r}} &= (-1)^{|\Sigma'|}\sum_{\sigma}\c_{\j_\sigma+\Sigma',\sigma}\cd_{\j_\sigma+\Sigma,\sigma}\ket{\j_\sigma, \Sigma} = (-1)^{|\Sigma'|}\hd_{\j_\sigma+\Sigma'} f_{\Sigma'}(\hat{s})\ket{0},
\end{align}
\end{subequations}
where~$|\Sigma|$ is the length of string~$\Sigma$, and~$\mathbf{r}=\pm \mathbf{e}_x,\pm\mathbf{e}_y$ is a unit step along the crystal lattice.
In the last step, we have defined the operator~$f_{\Sigma'}(\hat{s})\ket{0}$ that displaces the bosonic spinon background according to the string~$\Sigma$.
Now, we evaluate the hopping term~$\H_t$ between connected string states~$\ket{\j_\sigma,\Sigma_1}$ and~$\ket{\j_\sigma,\Sigma_2}$ with~$|\Sigma_2|-|\Sigma_1|=1$
\begin{align}
    \bra{\j_\sigma,\Sigma_2}\H_t\ket{\j_\sigma,\Sigma_1} = +t\cdot (-1)^{|\Sigma_1|+|\Sigma_2|}\bra{0} \h_{\j_\sigma+\Sigma_2} \hd_{\j_\sigma+\Sigma_2} [f_{\Sigma_2}(\hat{s})]^\dagger f_{\Sigma_1}(\hat{s}) \h_{\j_\sigma+\Sigma_1} \hd_{\j_\sigma+\Sigma_1}  \ket{0} = -t.
\end{align}
Therefore, the (sc) wavefunction amplitudes~$\psi_{\rm sc}(|\Sigma|)$ are positive and real for all~$|\Sigma|$ in this convention.
Note that we have neglected loop effects, which could lead to additional braiding of fermions. \\

\emph{Chargon-chargon states.}
Similarly, we define the (cc) basis states as
\begin{subequations}
\begin{align}
    \ket{\xc, \Sigma=\mathbf{r}} &= \c_{\xc,\sigma}\c_{\xc+\Sigma,\bar{\sigma}}\ket{0} = \hat{s}_{\xc,\sigma}\hat{s}_{\xc+\Sigma,\bar{\sigma}}\hd_{\xc}\hd_{\xc+\Sigma}\ket{0} \\
    \ket{\xc, \Sigma' = \Sigma+\mathbf{r}} &= (-1)^{|\Sigma'|+1}\sum_{\sigma}\c_{\xc+\Sigma',\sigma}\cd_{\xc+\Sigma,\sigma}\ket{\xc, \Sigma} = (-1)^{|\Sigma'|+1}\hd_{\xc}\hd_{\xc+\Sigma'} f_{\Sigma'}(\hat{s})\ket{0}.
\end{align}
\end{subequations}
In a tunneling process, the string length changes by~$|\Sigma_2|-|\Sigma_1|=1$ leading to amplitudes,
\begin{subequations}
\begin{align}
    \bra{\xc,\Sigma_2}\H_t\ket{\xc,\Sigma_1} &= +t\cdot (-1)^{|\Sigma_1|+|\Sigma_2|+2}\bra{0} \h_{\xc+\Sigma_2}\h_{\xc} \hd_{\xc+\Sigma_2} [f_{\Sigma_2}(\hat{s})]^\dagger f_{\Sigma_1}(\hat{s}) \h_{\xc+\Sigma_1} \hd_{\xc}\hd_{\xc, \Sigma_1}  \ket{0} = -t \\
    \bra{\xc+\mathbf{r},\Sigma_2}\H_t\ket{\xc,\Sigma_1} &= +t\cdot (-1)^{|\Sigma_1|+|\Sigma_2|+2}\bra{0} \h_{\xc+\Sigma_2}\h_{\xc+\mathbf{r}} \hd_{\xc+\mathbf{r}} [f_{\Sigma_2}(\hat{s})]^\dagger f_{\Sigma_1}(\hat{s}) \h_{\xc} \hd_{\xc}\hd_{\xc, \Sigma_1}  \ket{0} = -t
\end{align}
\end{subequations}
In this gauge choice, the (cc) wavefunction amplitudes~$\phi_{\rm cc}$ are real and positive.
\\

\emph{Open-closed channel coupling.}
Next, we consider states in the Hilbert space~$\mathscr{H} = \mathscr{H}_{\rm open}\oplus \mathscr{H}_{\rm closed}$ and determine the coupling matrix elements between an ${\rm (sc)}^2$ and (cc) state according to the above definitions. 
Let us consider string states that are connected via processes such as in Fig.~\ref{figRecombination}.
For the spin-flip processes, we find
\begin{align}
    \bra{\xc,\Sigma_{\rm cc}}\H_{J_\perp}\left( \ket{\j_\downarrow,\Sigma_\downarrow}\otimes \ket{\j_\uparrow,\Sigma_\uparrow} \right) =\frac{J_\perp}{2} \cdot (-1)^{|\Sigma_\uparrow|+|\Sigma_\downarrow|+|\Sigma_{\rm cc}| + 1} \bra{0} \h_{\xc+\Sigma_{\rm cc}}\h_{\xc} G_{\Sigma_{\rm cc},\Sigma_\downarrow,\Sigma_\uparrow}(\hat{s})   \hd_{\j_\downarrow+\Sigma_\downarrow}\hd_{\j_\uparrow+\Sigma_\uparrow}  \ket{0} = \frac{J_\perp}{2},
\end{align}
where the function~$G_{\Sigma_{\rm cc},\Sigma_\downarrow,\Sigma_\uparrow}(\hat{s})$ exchanges the spinons such that a (cc) string~$\Sigma_{\rm cc}$ is formed from the (sc) strings~$\Sigma_{\sigma}$.
Further, we have used that~$|\Sigma_{\rm cc}| =  |\Sigma_{\uparrow}| + |\Sigma_{\downarrow}| + \delta_{J_\perp}$ with~$\delta_{J_\perp} = -1,3$.
The case~$\delta_{J_\perp}=3$ corresponds to spin-flip terms as shown in Fig.~\ref{figRecombination}; the case $\delta_{J_\perp}=-1$ corresponds to a process where (sc)'s of length~$|\Sigma_1|=2$ and $|\Sigma_2|=0$ recombine into a (cc) with string~$|\Sigma_{\rm cc}|=1$.

Last, we need to evaluate the matrix elements for NNN tunneling processes~$t'$, which moves a chargon from site~$\j$ to site~$\j+\mathbf{r}'$.
To this end, we consider the matrix element with string states such as in Fig.~\ref{figRecombination},
\begin{align}
\begin{split}
    &\bra{\xc,\Sigma_{\rm cc}}\H_{t'}\left( \ket{\j_\downarrow,\Sigma_\downarrow}\otimes \ket{\j_\uparrow,\Sigma_\uparrow} \right) =\\
    & +t' \cdot (-1)^{|\Sigma_\uparrow|+|\Sigma_\downarrow|+|\Sigma_{\rm cc}| + 1} \bra{0} \h_{\xc+\Sigma_{\rm cc}}\h_{\xc} \hd_{\j_\sigma+\Sigma_\sigma+\mathbf{r}'}G_{\Sigma_{\rm cc},\Sigma_\downarrow,\Sigma_\uparrow}(\hat{s})\h_{\j_\sigma+\Sigma_\sigma} \hd_{\j_\downarrow+\Sigma_\downarrow}  \hd_{\j_\uparrow+\Sigma_\uparrow}  \ket{0}\\
    &= +t'.
\end{split}
\end{align}
Here, we used that only string states with~$|\Sigma_{\sigma}|=0$ and~$|\Sigma_{\bar{\sigma}}| \geq 0$ can be coupled to a (cc) string with length~$|\Sigma_{\rm cc}| = |\Sigma_{\sigma}| + |\Sigma_{\bar{\sigma}}| +1$.
Therefore, the above considerations resemble the results for hole doping in Eq.~\eqref{eq:hole-doublon-prefactors}. The doublon doping case is obtained analogously.

\end{document}